\documentclass[useAMS,usenatbib]{mn2e}
\usepackage{mn2e-breakabs}
\usepackage{graphicx}
\usepackage{subfigure}
\usepackage{times}
\usepackage{caption}
\usepackage{rotating}
\usepackage{rotfloat}
\usepackage{multicol}
\usepackage{array}
\usepackage{booktabs}
\usepackage{color}
\usepackage{afterpage}
\usepackage[multidot]{grffile}
\usepackage{bm}
\usepackage{float}
\usepackage{amsmath} 
\usepackage{amssymb}
\usepackage{hyperref}
\usepackage[normalem]{ulem}

\providecommand{\adsurl}[1]{\href{#1}{ADS}}

\newcommand{\hompc}{\,h\,{\rm Mpc}^{-1}}
\newcommand{\mpcoh}{\,h^{-1}\,{\rm Mpc}}
\newcommand{\Hunit}{\,{\rm km}\,{\rm s}^{-1}\,{\rm Mpc}^{-1}}
\newcommand{\mbi}{\boldsymbol}

\begin{document}
\title[Mock Comparison]
{
nIFTy Cosmology: Galaxy/halo mock catalogue comparison project on clustering statistics
}

\author[Chuang et al.]{
  \parbox{\textwidth}{
Chia-Hsun Chuang$^1$\thanks{E-mail: chia-hsun.chuang@uam.es, MultiDark Fellow},
Cheng Zhao$^2$,
Francisco Prada$^{1,3,4}$,
Emiliano Munari$^{5,10}$,
Santiago Avila$^{1,6}$,
Albert Izard$^7$,
Francisco-Shu Kitaura$^8$,
Marc Manera$^9$,
Pierluigi Monaco$^{5,10}$,
Steven Murray$^{11,12}$,
Alexander Knebe$^6$,
Claudia G. Sc\'occola$^{13,14}$,
Gustavo Yepes$^{6}$,
Juan Garcia-Bellido$^1$,
Felipe A. Mar\'in$^{15,16}$,
Volker M{\"u}ller$^8$,
Ramin Skibba$^{17}$,
Martin Crocce$^7$,
Pablo Fosalba$^7$,
Stefan Gottl{\"o}ber$^8$, 
Anatoly A. Klypin$^{18}$,
Chris Power$^{11,12}$,
Charling Tao$^{2,19}$,
Victor Turchaninov$^{20}$
}
  \vspace*{4pt} \\
$^1$ Instituto de F\'{\i}sica Te\'orica, (UAM/CSIC), Universidad Aut\'onoma de Madrid,  Cantoblanco, E-28049 Madrid, Spain \\ 
$^2$ Tsinghua Center for Astrophysics, Department of Physics, Tsinghua University, Haidian District, Beijing 100084, P. R. China\\ 
$^3$ Campus of International Excellence UAM+CSIC, Cantoblanco, E-28049 Madrid, Spain \\ 
$^4$ Instituto de Astrof\'{\i}sica de Andaluc\'{\i}a (CSIC), Glorieta de la Astronom\'{\i}a, E-18080 Granada, Spain \\ 
$^5$ Dipartimento di Fisica - Sezione di Astronomia, Universit\`a di Trieste, via Tiepolo 11, I- 34131 Trieste, Italy\\ 
$^6$ Departamento de F{\'i}sica Te{\'o}rica,  Universidad Aut{\'o}noma de Madrid, Cantoblanco, 28049, Madrid, Spain\\ 
$^7$ Institut de Ci\`encies de l'Espai, IEEC-CSIC, Campus UAB, Facultat de Ci\`encies, Torre C5 par-2, Barcelona 08193, Spain \\ 
$^8$ Leibniz-Institut f\"ur Astrophysik Potsdam (AIP), An der Sternwarte 16, D-14482 Potsdam, Germany\\ 
$^9$ University College London, Gower Street, London WC1E 6BT, UK\\ 
$^{10}$ INAF, Osservatorio Astronomico di Trieste, Via Tiepolo 11, I-34131 Trieste, Italy\\ 
$^{11}$ ICRAR, University of Western Australia, 35 Stirling Highway, Crawley, Western Australia 6009, Australia\\ 
 $^{12}$ ARC Centre of Excellence for All-Sky Astrophysics (CAASTRO)\\ 
$^{13}$ Facultad de Ciencias Astron\'omicas y Geof{\'\i}sicas - Universidad 
Nacional de La Plata. Paseo del Bosque S/N (1900). La Plata, Argentina \\ 
$^{14}$ CONICET, Rivadavia 1917, 1033, Buenos Aires, Argentina \\
$^{15}$ Technology, P.O. Box 218, Hawthorn, VIC 3122, Australia\\ 
$^{16}$ Centre for Astrophysics \& Supercomputing, Swinburne University of Technology, P.O. Box 218, Hawthorn, VIC 3122, Australia\\ 
$^{17}$ Department of Physics, Center for Astrophysics and Space Sciences, University of California, 9500 Gilman Drive, San Diego, CA 92093\\ 
$^{18}$ Astronomy Department, New Mexico State University, Las Cruces, NM, USA\\ 
$^{19}$ CPPM, Universit\'e Aix-Marseille, CNRS/IN2P3, Case 907, 13288 Marseille Cedex 9, France\\ 
$^{20}$ Keldysh Institute of Applied Mathematics, Russian Academy of Sciences, 125047 Moscow, Russia \\ 
}

\date{\today} 

\maketitle

\begin{abstract}
We present a comparison of major methodologies of fast generating
 mock halo or galaxy catalogues. The comparison is done for 
two-point (power spectrum and 2-point correlation function in real- and redshift-space), and the three-point clustering statistics (bispectrum and 3-point correlation function). 
The reference catalogues are drawn from the BigMultiDark $N$-body simulation.  Both friend-of-friends (including distinct halos only) and spherical overdensity (including distinct halos and subhalos) catalogues have been used with the typical number density of a large-volume galaxy surveys. We demonstrate that a proper biasing model is essential for reproducing the power spectrum at quasilinear and even smaller scales. With respect to various clustering statistics, a methodology based on perturbation theory and a realistic biasing model leads to very good agreement with N-body simulations. However,  for the quadrupole of the correlation function or the power spectrum, only the method based on semi-$N$-body simulation could reach high accuracy (1\% level) at small scales,
i.e., $r < 25 \mpcoh$ or $k > 0.15 \hompc$. 
Full $N$-body solutions will remain indispensable to produce reference catalogues. Nevertheless, we have demonstrated that the 
more efficient approximate solvers can reach a few percent accuracy in terms of clustering statistics at the scales interesting for the large-scale structure analysis.
This makes them useful for massive production aimed at covariance studies, to scan large parameter spaces, and to estimate uncertainties in data analysis techniques, such as baryon acoustic oscillation reconstruction, redshift distortion measurements, etc.
\end{abstract}


\section{Introduction}
The scope of galaxy redshift 
surveys has dramatically increased in the last years. The 2dF Galaxy Redshift Survey\footnote{http://www2.aao.gov.au/2dfgrs/} (2dFGRS) 
obtained 221,414 galaxy redshifts at $z<0.3$ \citep{Colless:2001gk,Colless:2003wz}, 
and the Sloan Digital Sky Survey\footnote{http://www.sdss.org} (SDSS, \citealt{York:2000gk}) collected 
930,000 galaxy spectra in the Seventh Data Release (DR7) at $z<0.5$ \citep{Abazajian:2008wr}.
WiggleZ\footnote{ http://wigglez.swin.edu.au/site/} collected spectra of 240,000 emission-line galaxies at $0.5<z<1$ over 
1,000 square degrees \citep{Drinkwater:2009sd, Parkinson:2012vd}, and the Baryon Oscillation Spectroscopic Survey\footnote{https://www.sdss3.org/surveys/boss.php} (BOSS, \citealt{Dawson:2012va}) of the SDSS-III project \citep{Eisenstein:2011sa} has surveyed 1.5 million luminous red galaxies (LRGs) at $0.1<z<0.7$ over 10,000 square degrees.
There are new upcoming ground-based and space experiments, such as
4MOST\footnote{http://www.4most.eu/} (4-metre Multi-Object Spectroscopic Telescope, \citealt{deJong:2012nj}), DES\footnote{http://www.darkenergysurvey.org} (Dark Energy Survey, \citealt{DES13}), DESI\footnote{http://desi.lbl.gov/} (Dark Energy Spectroscopic Instrument, \citealt{Schlegel:2011zz,Levi:2013gra}), eBOSS\footnote{http://www.sdss.org/sdss-surveys/eboss/} (Extended Baryon Oscillation Spectroscopic Survey), 
HETDEX\footnote{http://hetdex.org} (Hobby-Eberly Telescope Dark Energy Experiment, \citealt{Hill:2008mv}), 
J-PAS\footnote{http://j-pas.org} (Javalambre Physics of accelerating universe Astrophysical Survey, \citealt{Benitez:2014ibt}), 
LSST\footnote{http://www.lsst.org/lsst/} (Large Synoptic Survey Telescope, \citealt{Abell:2009aa}), 
Euclid\footnote{http://www.euclid-ec.org } \citep{Laureijs:2011gra}, 
and WFIRST\footnote{http://wfirst.gsfc.nasa.gov} (Wide-Field Infrared Survey Telescope, \citealt{Green:2012mj}), which would observe even larger galaxy samples.

Mock galaxy catalogues are essential for analysing the clustering signal drawn from these surveys. 
Tight constraints on cosmological models can be determined provided that the covariances of the clustering measurements are reliably estimated.
For such purpose, we need a large number of realizations of a simulation
designed to reproduce the volume of the Universe observed in a given
survey. $N$-body simulations are an ideal tool for
reproducing cosmological structures, e.g., LasDamas\footnote{http://lss.phy.vanderbilt.edu/lasdamas/} (Large Suite of Dark Matter Simulations), which has been used to analyse the SDSS-II galaxy sample (e.g., \citealt{Chuang:2010dv,Samushia:2011cs}), although running many realizations is expensive, or even unfeasible if such number has to be
very large (e.g., we might need $\sim 10^3$ or even more.). In order to
circumvent this problem, some alternatives have been proposed.
In the last decades, many new tools (see Table \ref{Table:mocks}) have been developed for reconstructing in an approximate way the large-scale structures down to the mildly non linear scales, allowing a fast generation of simulated volumes of the Universe. In this way, a direct computation of the covariance matrices by means of large numbers of realizations is possible.

\begin{table*}\normalsize
\centering
\resizebox{2\columnwidth}{!}{
\begin{tabular}{ll}
Methodology & reference\\
\hline
{\bf Log-Normal} & \citealt{Coles:1991if}\\
{\bf PTHalos}   &\citealt{Manera:2012sc,Manera:2014cpa}\\
{\bf PINOCCHIO}  (PINpointing Orbit-Crossing Collapsed Hierarchical Objects) & \citealt{Monaco:2001jg,Monaco:2013qta}\\
{\bf COLA}  (COmoving Lagrangian Acceleration simulation) & \citealt{Tassev:2013pn}\\
{\bf PATCHY} (PerturbAtion Theory Catalog generator of Halo and galaxY distributions) & \citealt{Kitaura:2013cwa,Kitaura:2014mja}\\
QPM (quick particle mesh) & \citealt{White:2013psd}\\
{\bf EZmock} (Effective Zel'dovich approximation mock catalogue) & \citealt{Chuang:2014vfa}\\
{\bf HALOgen} &   \citealt{Avila:2014aa}\\
\hline 
\end{tabular}
}
\caption{The methodologies of generating mock halo/galaxy catalogues developed in the last years. The methodologies included in this study are highlighted using bold font.}
\label{Table:mocks}
\end{table*}

In this paper, we compare these different methods, including COLA, EZmock, HALOgen, Log-Normal, PATCHY, PINOCCHIO, and PTHalos. We generate the halo mock catalouges using the same initial power spectrum (except Log-Normal model since it uses the observed correlation function as the input) and compare with the $N$-body simulation which also used the same initial power spectrum. This
comparison is meant to investigate the performances of the different
methods for computing the clustering properties (
power spectrum, correlation function,
bispectrum and three point correlation function) in real and redshift
space, leading to considerations on the capabilities of recovering the
properties of the baryonic acoustic oscillations (BAO) and redshift
space distortion.
We do not include the comparison of the positions of individual halos which can be provided by COLA, PINOCCHIO, and PTHalos. The other methods, i.e., EZmock, HALOgen, Log-Normal, and PATCHY, generate halos with some biasing models calibrated with the $N$-body simulations.

This paper -- emerging out of the 'nIFTy cosmology' workshop\footnote{http://popia.ft.uam.es/nIFTyCosmology} -- is organized as follows. In Section \ref{sec:bigmd}, we describe the reference $N$-body simulation catalogues used for our study. In Section \ref{sec:method}, we present a quick
description of the main characteristics of the different codes used in
this comparison work, highlighting their similarities and the differences. The results are presented in Section \ref{sec:result}, first for the main haloes and then also including the presence of substructures. We discuss the
results of the previous section, and finally conclude in Section \ref{sec:conclusion}.

\section{Reference $N$-body halo catalogues}
\label{sec:bigmd}
To test the different methods, we use a reference halo catalogue at redshift $z=0.5618$ extracted from the  BigMultiDark (BigMD) simulation\footnote{http://www.multidark.org/} \citep{Klypin:2014kpa}, which was performed using \textsc{gadget-2} \cite{Springel:2005mi} 
with $3840^3$ particles on a volume of $(2500$ $\mpcoh)^3$ assuming $\Lambda$CDM Planck cosmology with \{$\Omega_{\rm M}=0.307115, \Omega_{\rm b}=0.048206,\sigma_8=0.8288,n_s=0.96$\}, and a Hubble constant ($H_0=100\,h\Hunit$) given by  $h=0.6777$. 
Within the  MultiDark project a series of DM only simulations in different cosmologies and with different box sizes and resolutions have been performed (see \citealt{Klypin:2014kpa} for an overview). The MultiDark simulations have been used already to interpret the clustering of the BOSS galaxy sample \citep{Nuza:2012mw}. 

Haloes were defined based on two different algorithms. 
A friends-of-friends based code (called FoF; e.g., see \citealt{Riebe:2011gp}) and a spherical overdensity (SO) based code (called BDM; e.g., see \citealt{Klypin:1997sk,Riebe:2011gp}). The former code does not ab initio give subhaloes whereas the latter does, and haloes that are not subhaloes are also referred to as 'distinct haloes'. Note that we use 'BDM haloes' and 'SO haloes' interchangeably.
In this work, we use the FoF catalogue (linking length = 0.2) as our reference to compare between the different approximate methods; and also use the SO catalogues (obtained with BDM code) to discuss the effect of substructures. 
From the halo catalogue, we select a complete sample, selected by mass, with number density $3.5\times 10^{-4}$ $h^3\,{\rm Mpc}^{-3}$, which is similar to that of the BOSS galaxy sample at $z\sim0.5$. 
This abundance is equivalent to a mass cut of $\sim1\times10^{13}$ $M_{\odot}/h$ for the FoF catalogue and $\sim8.5\times10^{12}$ $M_{\odot}/h$ for the SO catalogue. 
Note that the BigMD simulation is designed to have the proper box size and mass resolution for constructing the mock galaxy catalogues for the BOSS survey which has collected the largest spectroscopic galaxy sample to date. 
While it would be interesting to go past these limits both in box size and mass resolution, we nevertheless leave this for future studies.

\begin{figure}
\centering
\includegraphics[width=0.5\textwidth]{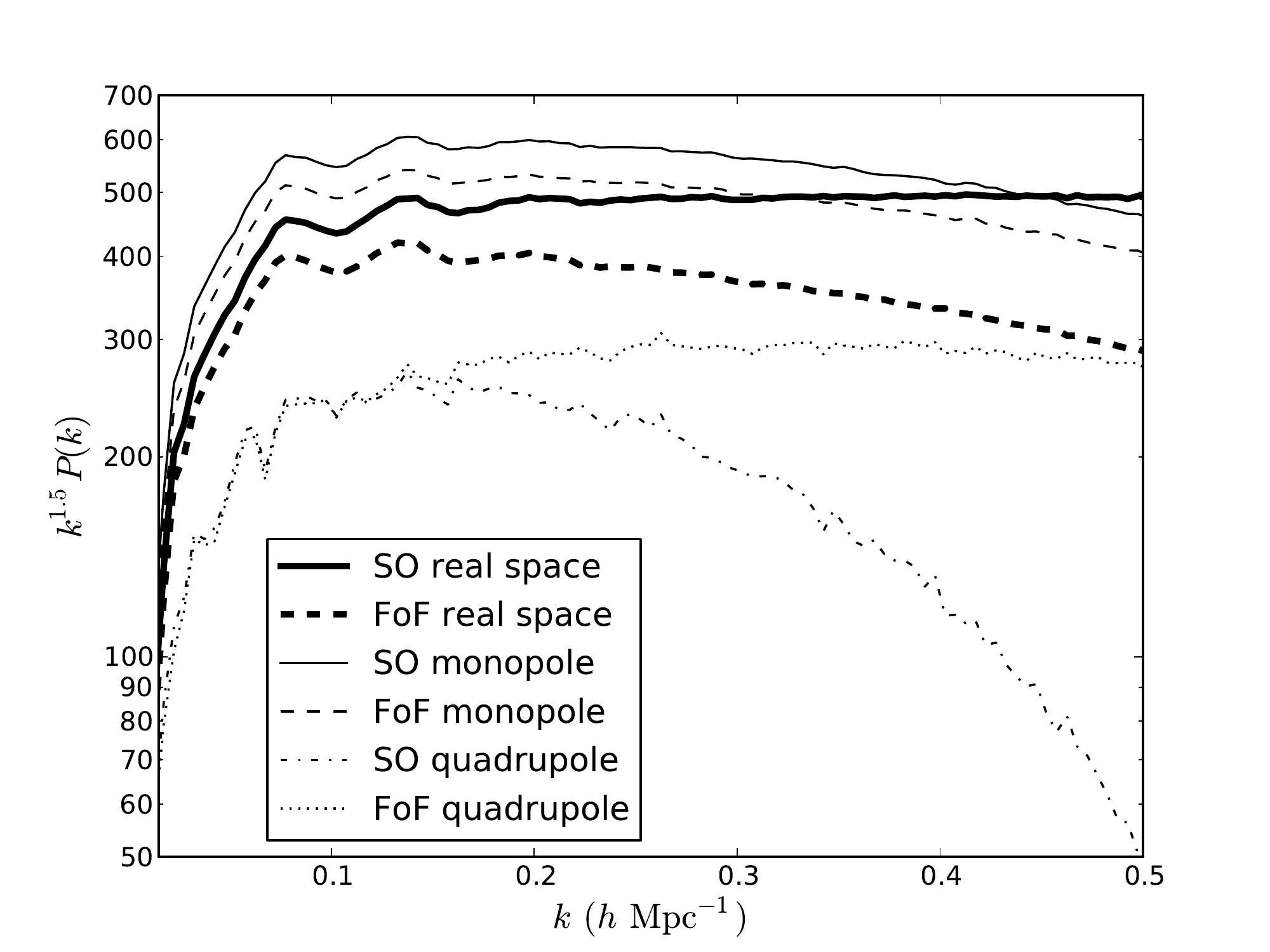}
\caption{Clustering statistics in real and redshift-space for the reference BigMultiDark SO and FoF catalogues, both with the same number density. Monopole of the power spectrum in real space for BigMD SO catalogue (thick solid line) and FoF catalogue (thick dashed line); monopole of power spectrum in redshift-space for SO (thin solid line) and FoF (thin dashed line); and quadrupole of  power spectrum in redshift-space for SO (dash-dotted line) and FoF (dotted line). In real space, SO monopole has higher amplitude due the clustering signal of subhaloes at small scales; but, in redshift space, the signal in the monopole is compensated out by the local motions. On the other hand, the quadrupole of the SO catalogue has much less signal due to the local motions.
}
\label{fig:bigmd_pk}
\end{figure}

Fig \ref{fig:bigmd_pk} displays the impact of substructures on the
large-scale clustering statistics. Specifically, we want to show
how the power spectrum at wavenumbers $k\lesssim 1\hompc$ is affected
by the one-halo term of the correlation function. Naively, one does not
expect that there is such an effect. After all, why should clustering at
$\lambda> 2\pi/k \sim 6\mpcoh$ be affected by inclusion of
subhaloes at much smaller scales? However, there are two effects.
The first one, is rather simple. There are more subhaloes of a given mass
(or circular velocity) in each massive distinct halo as compared with
less massive halo. When subhaloes are included, larger haloes give
proportionally larger contribution to the estimate of the power
spectrum. Because larger haloes are more biased, the power spectrum
(and the correlation function) are larger on all scales (see Fig 1). In practice,
this effect results in an almost scale-independent bias.
The second effect is more subtle: there is a change -- a boost due to
subhaloes -- in the power spectrum even when there is no change in the
large-scale correlation function. This happens because the power
spectrum and the correlation function are connected through an
integral relation. This effect results in a scale-dependent bias and
its effect gets progressively small for small wavenumbers $k$. In redshift-space, this effect on the monopole
is compensated due to the peculiar velocities, which yield to much smaller differences
between both BigMD catalogues: SO, including substructures, and FoF, which only contains distinct haloes (see Fig 1).
On the other hand, the quadrupole of the SO catalogue has much less signal due to those peculiar velocities.

\section{APPROXIMATE METHODS FOR MOCK COMPARISON}\label{sec:method}
\begin{figure}
  \includegraphics[width=\columnwidth]{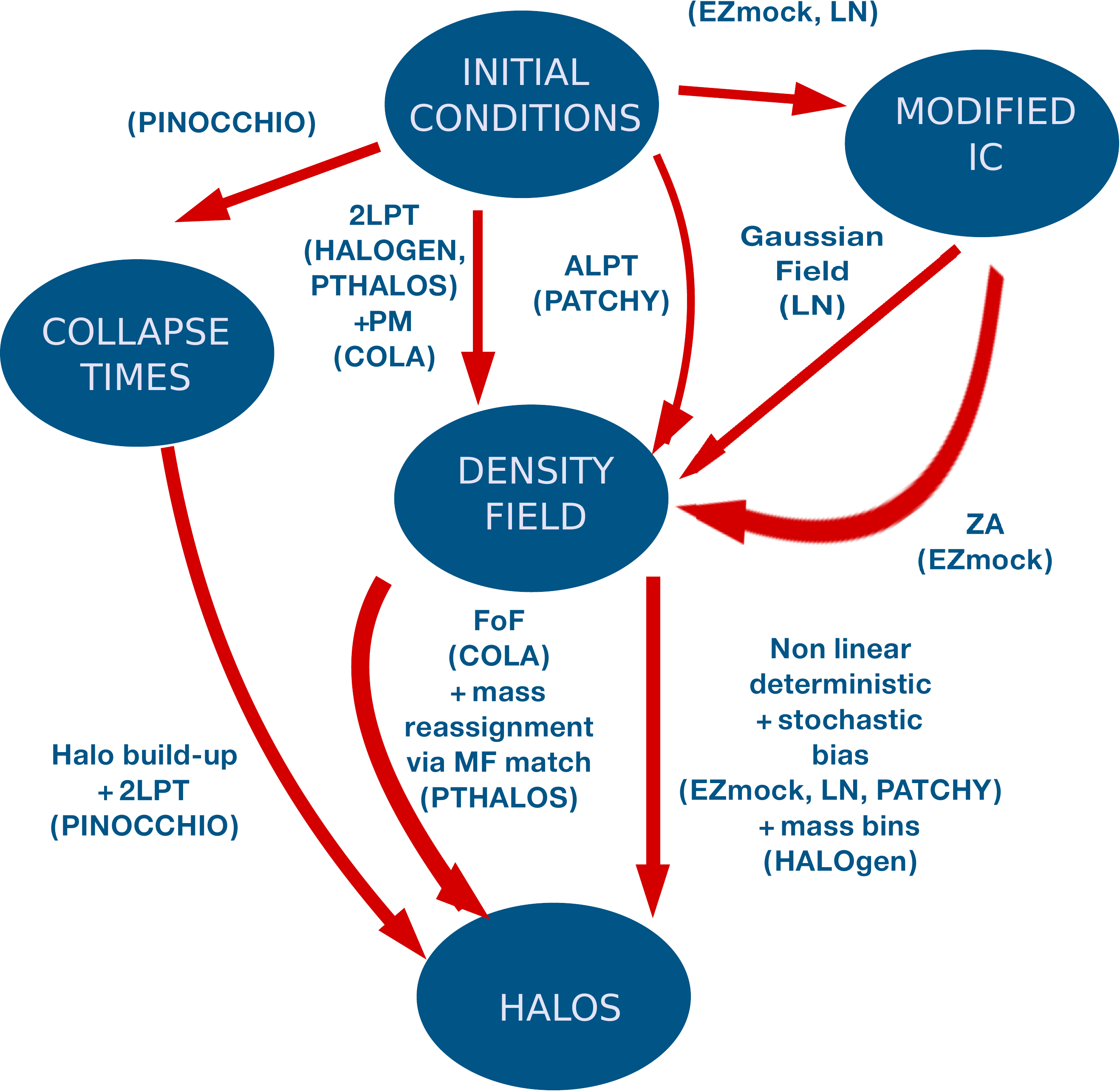}  
  \caption{A summary plot of different methodologies to generate mock halo catalogue. See context for detail description.}
 \label{Fig: methods_diagram} 
\end{figure}

The methods used for this comparison project start from a set of initial
conditions (ICs hereafter) with the aim of generating catalogues of dark matter
haloes. The way the different methods reach this goal can be divided
into three logical branches, as sketched in Fig \ref{Fig: methods_diagram}. PINOCCHIO reaches the first step by predicting the
collapse times of the particles from the ICs. The others instead
construct the density field before the identification or population of the
haloes. While most of them compute the density field directly from
the ICs, EZmock and LogNormal perform a modification of the initial conditions (see \citealt{Chuang:2014vfa}, and \citealt{Coles:1991if} for more details).

In Table \ref{Table: feature comparison} we compare the main technical
features of the methods. Below, we summarize the main ideas  and ingredients behind each method. For a detailed description of the methods we refer the
reader to the cited papers.

\begin{table*}\normalsize
\centering
\resizebox{2\columnwidth}{!}{
\begin{tabular}{c|c|c|c|c|c|c|c}
\hline
                                &	COLA	&	EZmock	&	HALOgen	&Log-Normal &	PATCHY 	&	PINOCCHIO	&	PTHalos	\\
\hline
Mass, Vel	                &	M + V	&	M(post-process) + V	&	M(binned) + V	& -- & M(post-process) + V	&	M+V	&	M +V			\\
Need of resolving the haloes	&	YES	&	NO	&	NO &	NO	&	NO	&	YES	&	YES 	\\
Initial conditions	        &	2LPTic	&	ZA	&	2LPTic	&Gaussian &	ALPT	&	N-GenIC; can read in graphic2	&	2LPTic		\\
Parallel	                &	MPI + openMP	&	openMP	&	openMP&	openMP	&	openMP	&	MPI + openMP	&	MPI	\\
Assumed MF	&	NO	&	YES	&	YES	&	-- & YES	&	NO	&	YES	\\
Assumed bias model	&	NO	&	YES	&	YES	& NO &	YES	&	NO	&	NO	\\
Substructures	&	Post-process	&	YES	&	Post-process	 &Yes&	YES	&	Post-process	&	Post-process	\\
Merger histories	&	NO	&	NO	&	NO	 &NO &	NO	&	YES	&	NO	\\
\# free params	&	0	&	7	&	1 (each mass bin)	& --&	7	&	5	&	1	\\
\# free params for z-space dist.	&	0	&	1	&	1	& -- &	2	&	0	&	0	\\
\# free params for MF	&	0	&	--	&	adopt MF	& -- &	--	&	5	&	adopt MF	\\
\# free params for bias	&	0	&	6	&	1	& -- &	5	&	0	&	0	\\
\hline 
\end{tabular}
}
\caption{Main technical features of the methodologies: COLA, PINOCCHIO, and PTHalos resolve haloes with some halo finders which can also provide the estimation of halo mass. EZmock and PATCHY provide halo catalogues with mass by applying a post-processing procedure (see Zhao et al. in prep.). The post-processing can be used to assign mass and other mass related quantities, e.g. circular veloctiy.
HALOgen constructs halo catalogues in mass bins; different initial condition codes are used to construct the dark matter density field for different methodologies; all the codes are using parallelisation techniques to speed up the computation; The methods using halo finders do not use bias models; EZmock, Log-Normal, and PATCHY construct the catalogues with substructures without post-processing; PINOCCHIO provides the merger histories. We also list the number of parameters used in each method. 
}
\label{Table: feature comparison}
\end{table*}

\subsection{COLA}
COLA \citep[COmoving Lagrangian Acceleration,][]{Tassev:2013pn} is a method to produce cheaper $N$-body simulations for large-scale structure. It uses a particle-mesh (PM) code with few timesteps to
solve for the residual displacements of particles with respect to
their trajectories calculated in Lagrangian Perturbation Theory
(LPT). Large scale dynamics is exactly solved since the 2LPT evolution
allows to recover the correct growth factor of fluctuations at such
scales. At small scales, the accuracy is controlled by the number of
timesteps \citep[in][they propose 10 timesteps starting at redshift
9]{Tassev:2013pn}.

The key point of this method is how the equation of motion is
rewritten. The displacement field is decomposed in two terms, one
describing the 2LPT trajectory and another one for the residual
displacement:
\begin{equation}
x_{res}\equiv x-x_{LPT},
\end{equation}
so that the equation of motion schematically reads
\begin{equation}
\partial_{t}^{2}x_{res}=-\nabla\Phi-\partial_{t}^{2}x_{LPT}.
\end{equation}
COLA discretizes the time derivatives only on the left-hand side,
while uses the LPT expression at the right-hand side.

In \cite{Tassev:2013pn} they developed a serial code for the
demonstration of the method. Afterwards, J. Koda parallelized it and made
it suitable for running large ensembles of simulations, as done
in \cite{Kazin:2014qga}. 
For an optimized and parallel version of COLA, including lightcone outputs, see Izard et al. 2014 in preparation.

\subsection{EZmock}
EZmock \citep[Effective Zel'dovich approximation mock catalogue,][]{Chuang:2014vfa} is constructed from the Zel'dovich approximation density field.
It absorbs the nonlinear effect and halo bias (i.e. linear, nonlinear,
deterministic, and stochastic bias) into some effective modeling with
few parameters, which can be efficiently calibrated with $N$-body
simulations.
The following required steps are recursively applied until convergence:
\begin{enumerate}
 \item[(I)] generation of the dark matter density field on a grid using the Zel'dovich approximation (ZA);
\item[(II)] mapping the probability distribution function (PDF) of haloes measured in BigMD to the ZA density field;
\item[(III)] adding scatter to the PDF mapping scheme by 
\begin{equation}
\rho_{s}(\boldsymbol r) = 
\left\{ \begin{array}{ll}
         \rho_{o}(\boldsymbol r)(1+G(\lambda)) & \mbox{if $G(\lambda) \ge 0$};\\
        \rho_{o}(\boldsymbol r)\exp(G(\lambda)) & \mbox{if $G(\lambda) < 0$},
\end{array} \right.
\end{equation}
where $\rho_s(\boldsymbol r)$ and $\rho_o(\boldsymbol r)$ are the ZA density field after and before the scattering respectively. $G(\lambda)$ is a random number drawn from the Gaussian distribution with width $\lambda$. The exponential function is used to avoid the negative density; 
\item[(IV)] fitting the amplitude of the power spectrum and bispectrum with a density threshold and saturation before the scattering scheme by
\begin{equation}
\rho_{o'}(\boldsymbol r)=
\left\{ \begin{array}{ll}
  0, & \mbox{if $\rho_{o}(\boldsymbol r) < \rho_{\rm th}^{\rm low}$}; \\
  \rho_{\rm th}^{\rm high}, & \mbox{if $\rho_{o}(\boldsymbol r) > \rho_{\rm th}^{\rm high}$},
\end{array} \right.
\end{equation}
where $\rho_{o'}(\boldsymbol r)$ is the modified density, $\rho_{o}(\boldsymbol r)$ is the original ZA density, $\rho_{\rm th}^{\rm low}$ and $\rho_{\rm th}^{\rm high}$ are the density threshold and density saturation respectively;
\item[(V)] fitting the shape of the final power spectrum by modifying the tilt in the initial input power spectrum with a scale-dependent function by
\begin{equation}
P_{\rm ePK}(k)=P_{\rm eBAO}(k)\cdot(1+Ak),
\end{equation}
where $A$ is a free parameter;
\item[(VI)] fitting baryon acoustic oscillations (BAOs) by enhancing the amplitude of BAOs in the initial input power spectrum by
\begin{equation}
P_{\rm eBAO}(k)=(P_{\rm lin}(k)-P_{\rm nw}(k))\exp(k^2/k_*^2) + P_{\rm nw}(k),
\end{equation}
where $P_{\rm eBAO}(k)$ is the BAO enhanced power spectrum, $P_{\rm lin}(k)$ is the linear power spectrum, $P_{\rm nw}(k)$ is the smoothed no-wiggle power spectrum obtained by applying a cubic spline fit to $P_{\rm lin}(k)$, and $k_*$ is usually known as the damping factor (however, for the damping model, one should use $\exp(-k^2/k_*^2)$ instead);
\item[(VII)] computing the peculiar motions $v$ within the ZA for each object
 by adding to the linear coherent motion, which is proportional to the ZA displacement field, a dispersion term modeled by a random Gaussian distribution, i.e., 
\begin{equation}
v_i(\boldsymbol r)=B\psi_i(\boldsymbol r)+G(\lambda'),
\end{equation}
where $B$ is a constant corresponding to linear growth; $\psi$ is the displacement field, $i$ denotes the direction $x$, $y$, or $z$; and $G(\lambda)$ is a random number drawn from the Gaussian distribution with width $\lambda'$.
\end{enumerate}

\subsection{HALOGEN} \label{sec:halogen}
The aim of HALOgen \citep{Avila:2014aa} is to provide a simple and efficient approximate method for generating mock halo catalogues with correct mass-dependent 2-point statistics. 
The basic algorithm is as follows:
\begin{itemize}
	\item[(I)] Create a cosmological matter field, sampled by $N$ particles using 2LPT.
	\item[(II)] Sample a number of halo masses corresponding to the desired number density from an appropriate analytical mass function (or reference $N$-body simulation).
	\item[(III)] Reconstruct the density field on a regular grid of cell-size $\l_{cell} \approx 2d_{\rm part}$ (twice the mean-interparticle distance, for this comparison we used $l_{cell} = 4\mpcoh$).
	\item[(IV)] Distribute haloes into mass bins (for this comparison we use 8 bins), and for each bin $M_j$ from highest to lowest mass, place each halo in the following way:
			\begin{itemize}
				\item Choose a cell with probability $P(i|M_j) \propto \rho_i^{\alpha(M_j)}$.
				\item Place the halo on a random 2LPT particle within the cell.
				\item Ensure the halo does not overlap previous halo centres (if so, repeat the cell choice).
				\item Decrease the mass of the cell by the mass of the halo (ensuring mass conservation on scales of $l_{cell}$).
			\end{itemize}	
	\item[(V)] Assign particle velocities to haloes with a factor $\vec{v}_{\rm h}=f_{\rm vel}\cdot  \vec{v}_{\rm p}$, computed as the ratio of the velocity dispersions of the selected particles and the reference halo catalog:
 $f_{\rm vel}=\frac{\sigma(\vec{v}_{\rm p})}{\sigma(\vec{v}_{\rm ref})}$ 
\end{itemize}
The only free parameter of the placement is $\alpha(M)$, which primarily controls the linear halo bias. It can be fitted once for a given cosmology, redshift and $l_{cell}$, and used for any number of random initial conditions. An additional parameter controls the velocity bias, and is simply calculated via the ratio of the variance of the $N$-body velocities to the 2LPT particle velocities.
The efficiency of HALOgen is primarily constrained by the 2LPT step, as the algorithms intrinsic to HALOgen are very fast.

\subsection{Log-Normal}
The distribution of galaxies on intermediate to large scales ($> 10\mpcoh$) has been found to follow a lognormal distribution
\citep[see][]{hubble1934,Wild:2004me} especially when correcting for shot
noise effects (see \citealt{Kitaura:2009uh}). The physical argument for this
behaviour has been found in the continuity equation, as the co-moving
solution of the evolved density field is related to the linear density
field through a logarithmic transformation when shell-crossing is
neglected \citep[see][]{Coles:1991if,Kitaura:2011ay}. This implies that
under the assumption of Gaussian primordial fluctuations the evolved
density field is expected to be lognormal distributed on intermediate
to large scales. It has the advantage that its two-point statistics
can be exactly controlled. Therefore, it has been widely used to study cosmic variance (and covariance matrices) in large-scale structure measurements (e.g., \citealt{Cole:2005sx,Percival:2009xn,Reid:2009xm,Blake:2011wn,Beutler:2011hx}).
The Log-Normal mock is constructed with the following steps:
\begin{itemize}
 \item[(I)] Given a input correlation function,
$\xi(r)$, the Gaussian field correlation function is obtained by
\begin{equation}
\xi_G(r)=\ln[1+\xi(r)],
\end{equation}
and this can be Fourier transformed to the power spectrum, $P_G(k)$, 
\item[(II)] A Gaussian density field $\delta_G(r)$ is generated on the grid with the power spectrum, $P_G(k)$, 
\item[(III)] A lognormal field is calculated by
\begin{equation}
\delta_{LN}(r)=\exp\left[\delta_G(r)-\frac{\sigma_G^2}{2}\right]-1,
\end{equation}
where $1+\delta_{LN}(r)$ is the lognormal density field which is always positive by definition and $\sigma_G^2$ is the variance of the  Gaussian density field which
can be calculated by
\begin{equation}
 \sigma_G^2=\sum^{N_{grid}}_{i,j,l=1}P_G\left[(k^2_{x_i}+k^2_{y_j}
 +k^2_{z_l})^\frac{1}{2}\right],
\end{equation}
where $N_{grid}$ is the number of grid points, $k_{m_n}=\frac{2\pi}{L}\left(n-\frac{N_{grid}}{2}\right)$, $L$ is the box length, and $m=x$, $y$, or $z$. 
\item[(IV)] Draw the Poisson random variables with the
means given by this lognormal field.
\end{itemize}
In principle, one could assign the velocity to the Log-Normal mocks (e.g., see \citealt{White:2013psd}), but it is not done in this study.

\subsection{PATCHY}
PATCHY \citep{Kitaura:2013cwa} relies on modeling the large-scale
structure density field with an efficient approximate gravity solver
and populating the density field following a non-linear, scale-dependent, and
stochastic biasing description.
Below, the main ingredients are listed:

\begin{enumerate}
\item[(I)] A one-step gravity solver based on augmented Lagrangian
perturbation theory (ALPT, \citealt{Kitaura:2012tj}), correcting
second order LPT (2LPT) in the high and low density regimes with a nonlinear local term derived from the spherical collapse model (SC) matching $N$-body
simulations.  
In this approximation the displacement field  $\mbi\Psi_{\rm ALPT}(\mbi q,z)$, mapping a distribution of dark matter particles at initial Lagrangian positions $q$ to the final Eulerian positions $\mbi x(z)$ at redshift $z$ ($\mbi x(z)=\mbi q+\mbi\Psi(\mbi q,z)$), is split into a long-range and a short-range component, given by 2LPT and SC, respectively:
\begin{equation}\label{eq:disp}
\mbi\Psi_{\rm ALPT}(\mbi q,z)={\cal K}(\mbi q,r_{\rm S}) \circ \mbi\Psi_{\rm 2LPT}(\mbi q,z)+\left(1-{\cal K}(\mbi q,r_{\rm S}) \right)\circ \mbi\Psi_{\rm SC}(\mbi q,z)\;
\end{equation}
\item[(II)] A deterministic bias model relating the expected number density of haloes $\rho_h$ to the dark matter density field $\rho_{\rm M}$ including a thresholding $\rho_{\rm th}$
and (or) an exponential cut-off $\exp\left[-\left(\frac{\rho_{\rm M}}{\rho_\epsilon}\right)^{\epsilon}\right]$, a power law density relation $\rho_{\rm M}^\alpha$:
\begin{equation}
\rho_h=f_h\,\theta(\rho_{\rm M}-\rho_{\rm th})\,\rho_{\rm M}^\alpha\,\exp\left[-\left(\frac{\rho_{\rm M}}{\rho_\epsilon}\right)^{\epsilon}\right]{,}
\end{equation}
with 
\begin{equation}
\label{eq:numden}
f_h=\bar{N}_h/\langle\theta(\rho_{\rm M}-\rho_{\rm th})\,\rho_{\rm M}^\alpha\,\exp\left[-\left(\frac{\rho_{\rm M}}{\rho_\epsilon}\right)^{\epsilon}\right]\rangle{,}
\end{equation}
 and \{$\rho_{\rm th},\alpha,\epsilon,\rho_\epsilon$\} the parameters of the model;
\item[(III)] A sampling step, which deviates from Poissonity modelling over-dispersion and stochasticity in the bias relation, in particular using the negative binomial distribution function:
\begin{equation}
P(N_i\mid\rho_{hi},\beta)=\frac{\lambda_i^{N_i}}{N_i!}\frac{\Gamma(\beta+N_i)}{\Gamma(\beta)(\beta+\rho_{h})^{N_i}}\frac{1}{(1+\rho_{h}/\beta)^\beta} \;
\end{equation}
with $\beta$ being the stochastic bias parameter.
\item[(IV)] The parameters are constrained to efficiently match the halo (or
galaxy) probability distribution function (PDF) and the power spectrum
for a given number density. In this way we can match the 3-point
statistics,
\item[(V)] Peculiar velocities are split into a coherent and a
quasi-virialised component. The coherent flow is obtained from ALPT
and the dispersion term is sampled from a Gaussian distribution
assuming a power law relation with the local density.
\end{enumerate}

\subsection{PINOCCHIO}
PINOCCHIO\footnote{http://adlibitum.oats.inaf.it/monaco/Homepage/Pinocchio/index.html} \cite{Monaco:2001jg,Monaco:2013qta} is based on the
ellipsoidal collapse, solved with the aid of 3rd-order LPT, to compute
the time at which mass elements collapse (in the orbit-crossing
sense), and Extended Press \& Schecther (EPS) to deal with multiple
smoothing radii. It starts from the generation of a linear density
field on a regular grid in Lagrangian space, in the same way as
initial conditions are generated for an $N$-body simulation. The
density field is smoothed on a set of scales, and the collapse time is
computed for each particle and at each smoothing radius. The earliest
time is recorded as the bona-fide estimate of collapse time.

The collapsed medium is then fragmented into disjoint haloes by
applying an algorithm that mimics the hierarchical formation and
merging of haloes. This works as follows: particles are sorted in order
of increasing collapse times. When a particle collapses, the fate of
its six Lagrangian neighbours is checked. If all neighbours have not
collapsed, then a new group with one particle is formed. If one
neighbour already belongs to a group, then the particle and the group
are displaced from the Lagrangian to the Eulerian space using
Zel'dovich or 2LPT displacements computed at the same time of collapse
of the particle. If the particle gets within the ``virial radius'' of
the group, then it is accreted to the group, otherwise it is tagged as
a ``filament'' particle. Filaments are later accreted on a group each
time a neighbouring particle is accreted on the same group. If the
Lagrangian neighbours of the collapsing particle belong to more
groups, then the groups are displaced to check whether the center of
mass of one group gets within the ``virial radius'' of the other. If
this takes place the two groups are merged. The estimate of the
``virial radius'' implies the use of parameters, as fully explained in
\cite{Monaco:2001jg}. These parameters are chosen requiring to
reproduce a given (universal) mass function. Their values are
independent of redshift, mass resolution and cosmology, so once they
are fixed the code can be applied to any configuration.

Because of the algorithm used to create haloes, PINOCCHIO can also generate
accurate merger histories of haloes with continuous time sampling.

In this paper we use a new version of the code, with 2LPT displacements and a better tuning of the mass function, that will be presented in a forthcoming paper.
To compute the covariance of 2-point correlation function for the VIPERS survey \citep{delaTorre:2013rpa} used a limited set of lightcones drawn from one of the MultiDark simulations and 200 mocks constructed with the PINOCCHIO code described above, using the Shrinkage technique \citep{Pope:2007vz} to deal with the bias introduced by the approximate code.

\subsection{PTHalos}
The basic steps in this method, inspired by \cite{Scoccimarro:2001cj}, can be summarised as follows \citep{Manera:2012sc,Manera:2014cpa}:

(I) Create a dark matter particle field based 2LPT.

(II) Identify halos using a Friends-of-Friends (FoF, \citealt{Davis:1985rj}) halo finder with an appropriately chosen linking length. Alternatively, one can identify halos with Spherical Overdensity with an equivalent density threshold. 

(III) The halos can be later populated with galaxies. 

Because the 2LPT dynamics is an approximation to the true dynamics of the dark matter field, it yields halo densities that consistently differ from the $N$-body densities. Consequently, the FoF linking length of the 2LPT matter field, $b_{2LPT}$, needs to be rescaled from the value used in $N$-body simulations, $b_{sim}$. The rescaling is given by

\begin{equation}
b_{2LPT} = b_{sim} \left( \frac{\Delta_{vir}^{sim}}{\Delta_{vir}^{2LPT}} \right)^{(1/3)} \; .
\label{linkscaled}
\end{equation}

Both the halo virial overdensity in $N$ -body simulations, $\Delta_{vir}^{sim}$, and its corresponding value in the 2LPT field, $\Delta_{vir}^{2LPT}$  are easy compute. For the $N$-body case we take the value of \cite{Bryan:1997dn},
\begin{equation}
\Delta_{vir}^{sim}=(18\pi^2+82(\Omega_m(z)-1)-39(\Omega_m(z)-1)^2) /\Omega_m(z)\; ,
\label{BryanDvir}
\end{equation}
where
\begin{equation}
\Omega_m(z)=\Omega_m (1+z)^3/H^2(z).
\end{equation}

For the Lagrangian case, $\Delta_{vir}^{2LPT}$ can be obtained from the relation between the Lagrangian and Eulerian coordinates, giving a value, within the spherical collapse approximation, of ~35.4 times the mean background density \citep{Manera:2012sc}. 

\cite{Scoccimarro:2001cj} originally constructed a merger tree to assign halos masses in cells. This method adopts a mass function and imparts it to the rank-ordered halos found by the halo finder. PTHalos has been used for BOSS galaxy clustering analysis \citep{Manera:2012sc,Manera:2014cpa}.


\section{results}\label{sec:result}
In this section, we present and compare the performance of all the methodologies to generate halo catalogues including FoF catalogue (distinct haloes only) and SO catalog (distinct and subhaloes) described in the previous sections. Table \ref{Table:setting} lists the particle mesh sizes adopted by the different methodologies, and also shows whether the reduced white noise is used. 
Note that the mesh sizes used by these methodologies are different from the BigMD simulation ($3840^3$), so that we cannot use the white noise used by the BigMD as initial condition directly. We compute the reduced white noise by averaging and rescaling the noise on the neighbor grid points to have the white noise on the smaller mesh size. The reduced white noise will share part but not the whole of the noise with the BigMD simulation. 
One should keep in mind that the adopted mesh serves different purposes for the different codes and also affects the timing and required resources. For some methodologies, the mesh size influences the scales on which haloes are resolved whereas other methodologies use the reference catalogue to calibrate their specific biasing model to arrive at the final mock halo catalogue.

\begin{table*} \normalsize
\centering
\resizebox{2\columnwidth}{!}{
\begin{tabular}{lp{2.5cm}| p{2.5cm} |p{2.5cm}|p{2.5cm}|p{2.5cm}|p{2.5cm}|p{2.5cm}|p{2.5cm}|p{1cm}|}
\hline
                                &BigMD&	COLA	&	EZmock	&	HALOGEN	& Log-normal &	PATCHY	&	PINOCCHIO	&	PTHalos	 \\
\hline
Particle mesh size	            &$3840^3$    &	$1280^3$&	$960^3$	&	$1280^3$	& $1280^3$ &	$960^3$	&	$1920^3$	&	$1280^3$	 \\
 & & ($3840^3$ for force)	& & & & & & \\
Using white noise	        &YES &	NO	&	YES	&	YES	&  NO &	YES	&	YES	&	NO		\\
CPU-hour &800,000 & 130 & 1.3 & 6.7 & 0.5 & 8 & 440 & 45 \\
Memory &$8$Tb &550Gb &28Gb & 130Gb & 15Gb & 24Gb & 890Gb & 112Gb \\
\hline 
\end{tabular}
}
\caption{This table lists the particle mesh sizes adopted by the different approximate methods presented in this comparison project; whether the reduced white noise is used; and the computational requirements including CPU-hours and memory used for the mocks provided in the study.
Although using the BigMultiDark white noise is not required for mock generation, it will have an effect on the performances at large scales.
Note that the computational requirements might depend on the machines used which could be a factor of two or even more.
}
\label{Table:setting}
\end{table*}

\subsection{Mocks for FoF catalogues}
Here, we compare the different mocks with the BigMultiDark FoF reference catalog (see Sec. \ref{sec:bigmd}).
The mesh size used for computing the statistics is $960^3$ if applicable. 

\begin{figure}
\centering
\includegraphics[width=0.5\textwidth]{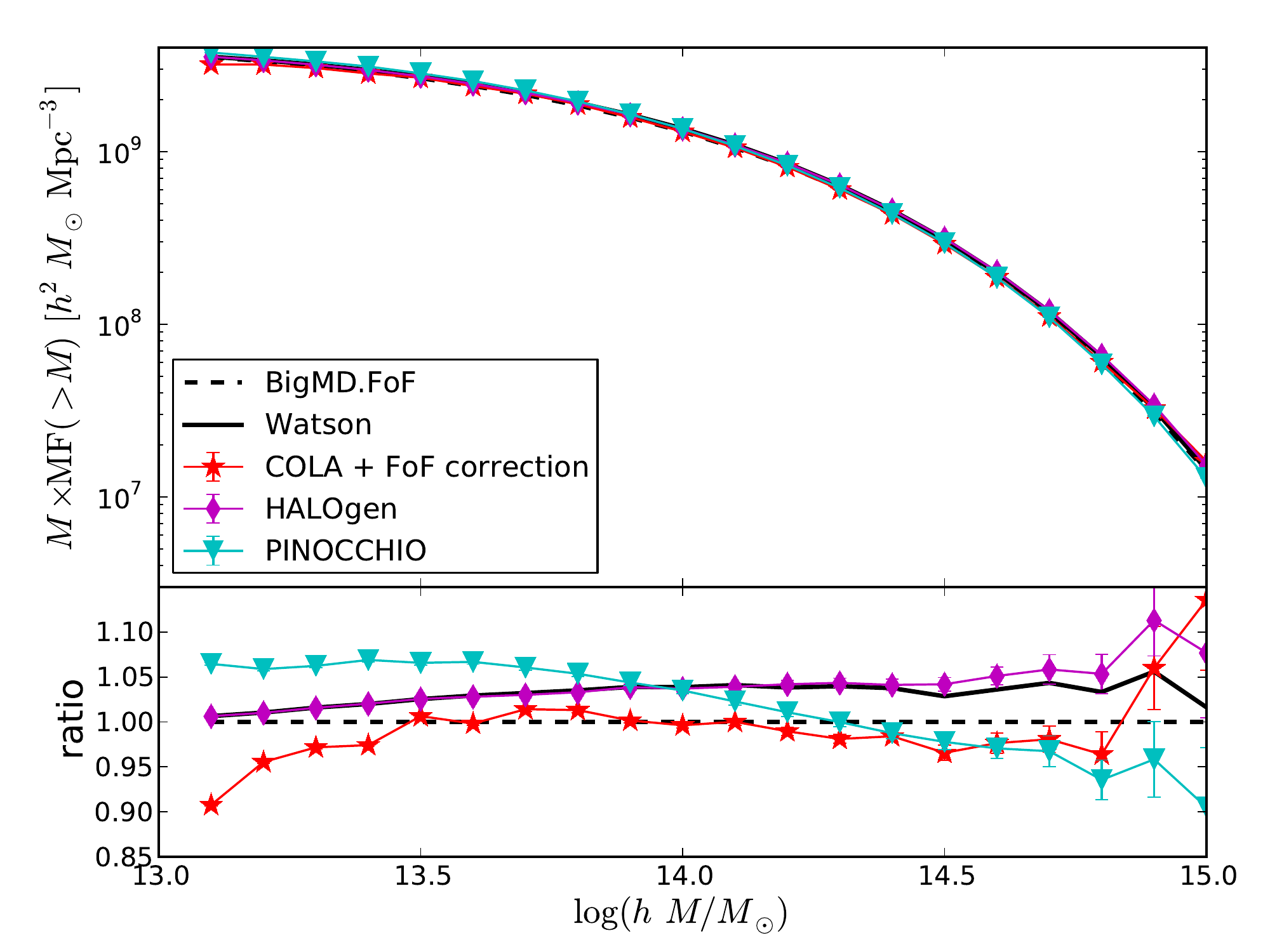}
\caption{Cumulative mass functions comparing with the BigMultiDark FoF reference catalog. The error-bars were estimated using Jack-knife resampling using 64 different sub-volumes. All the methods reproduce the numerical mass function to 5\% accuracy.}
\label{fig:mass_func}
\end{figure}

Some of the methods provide the masses for the halo catalog. Fig. \ref{fig:mass_func} shows the mass functions provided by COLA, HALOgen, and PINOCCHIO, compared with that from the BigMultiDark FoF catalogue. COLA FoF masses include the Warren correction due to
discrete halo sampling \citep{Warren:2005ey}:
\begin{equation}
M = N m_p (1-N^{-0.6}),
\end{equation}
where N is the number of particles in the halo and $m_p$ is the particle mass. 
HALOgen uses a theoretical mass function from \cite{Watson:2012mt}
as an input. All the methods reproduce the numerical mass function to 5\% accuracy.
The other mocks which do not provide masses could be assigned with a post-processing based on the particle density field (see \citealt{Zhao:2015jga}).

\subsubsection{2-point clustering statistics of FoF catalogues}
\begin{figure}
\centering
\includegraphics[width=0.5\textwidth]{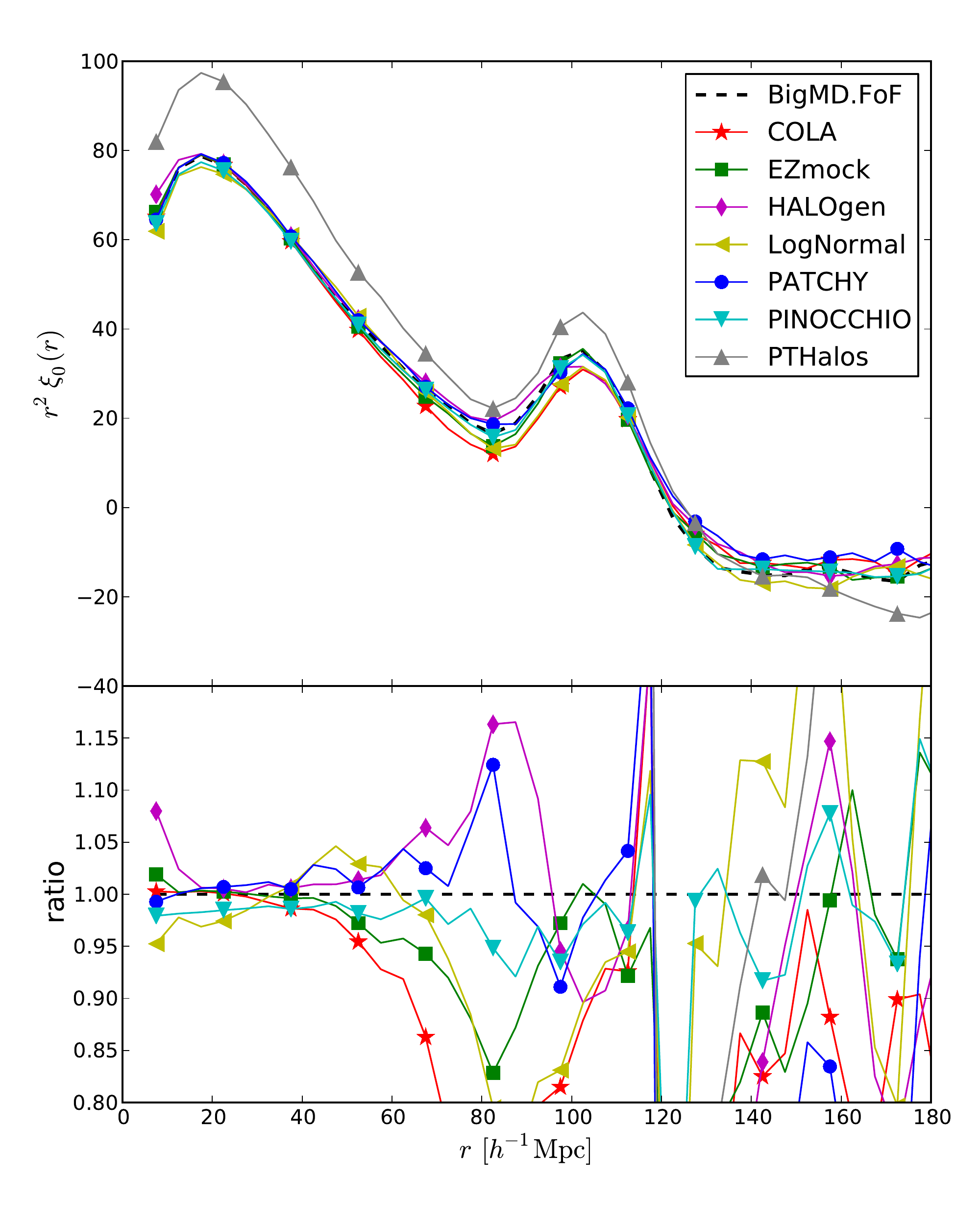}
\caption{Comparison of the monopole of the correlation function in real-space. Dashed line corresponds to the BigMultiDark FoF reference catalogue.
COLA FoF masses include the correction due to
discrete halo sampling \citep{Warren:2005ey}.
}
\label{fig:FOF_cf_r}
\end{figure}

\begin{figure}
\begin{center}
 \subfigure{\label{fig:cf_z_mono}\includegraphics[width=0.50 \textwidth]{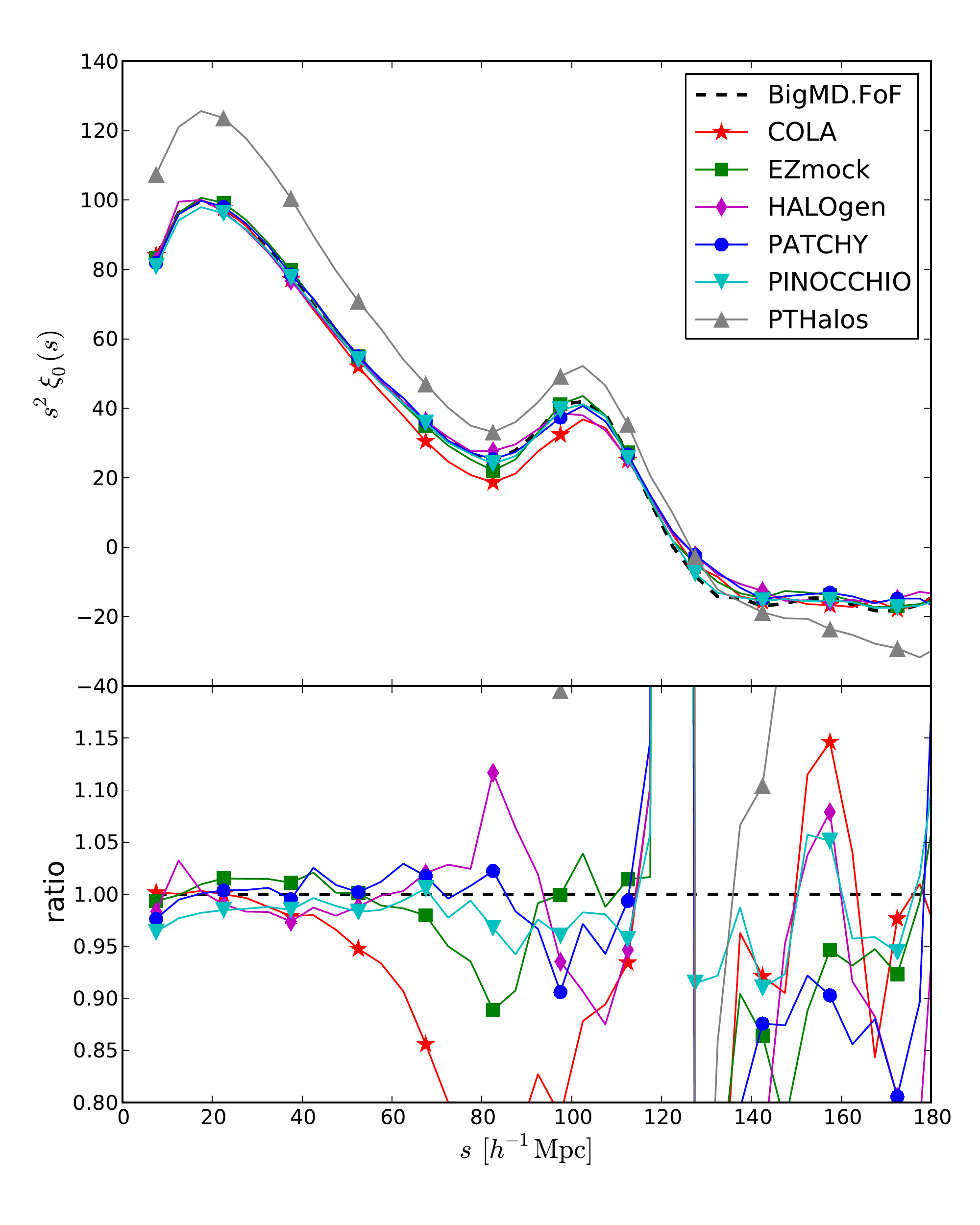}}
 \subfigure{\label{fig:cf_z_quad}\includegraphics[width=0.50 \textwidth]{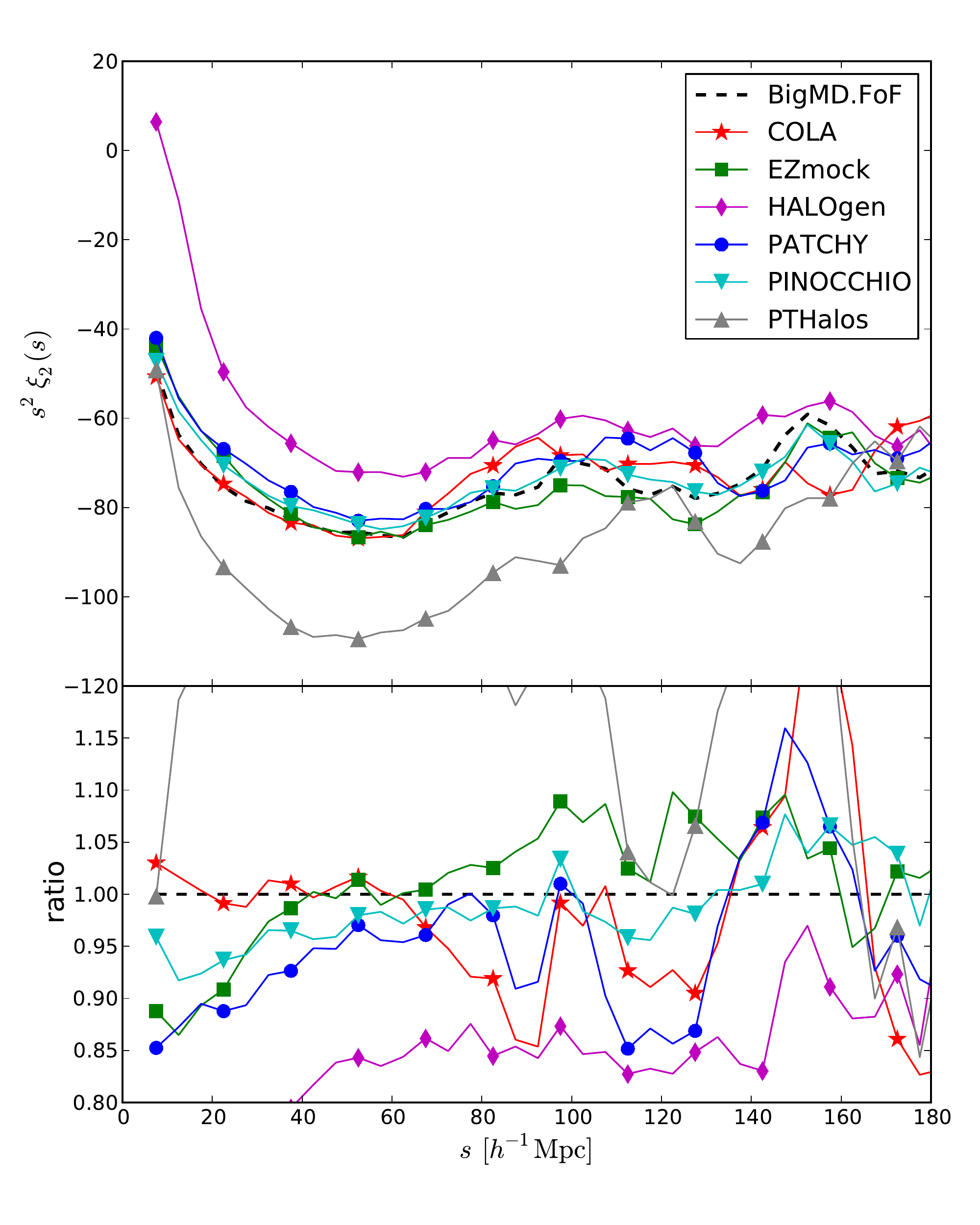}}
\end{center}
\caption{Top panel: comparison of the monopole of the correlation function in redshift space. Bottom panel: performance results for the quadrupole of the correlation function in redshift-space. Dashed lines correspond to the BigMultiDark FoF reference catalogue.}
\label{fig:FOF_cf_z}
\end{figure}

2-point clustering statistics is one of the most useful measurements in the clustering analysis of the galaxy surveys. Fig. \ref{fig:FOF_cf_r} shows the monopole of the correlation function in real-space.
Besides PTHalos, all the mocks agree with the simulation within $5\%$ at the scales between 10 to 50 $\mpcoh$. At larger scales, the deviations are basically due to noise. 
Fig. \ref{fig:FOF_cf_z} shows the monopole and quadrupole of the correlation function in redshift-space.
The comparison of the monopole in redshift space is basically the same as in real space. We have checked that the deviations that COLA has at large scales are mainly due to sample variance (COLA did not use the BigMultiDark white noise). For the quadrupole, COLA agrees with the BigMultiDark within $5\%$ down to the minimum scale we measured (10 $\mpcoh$); PINOCCHIO agrees within $10\%$; EZmock and PATCHY are within $15\%$.

\begin{figure}
\centering
\includegraphics[width=0.5\textwidth]{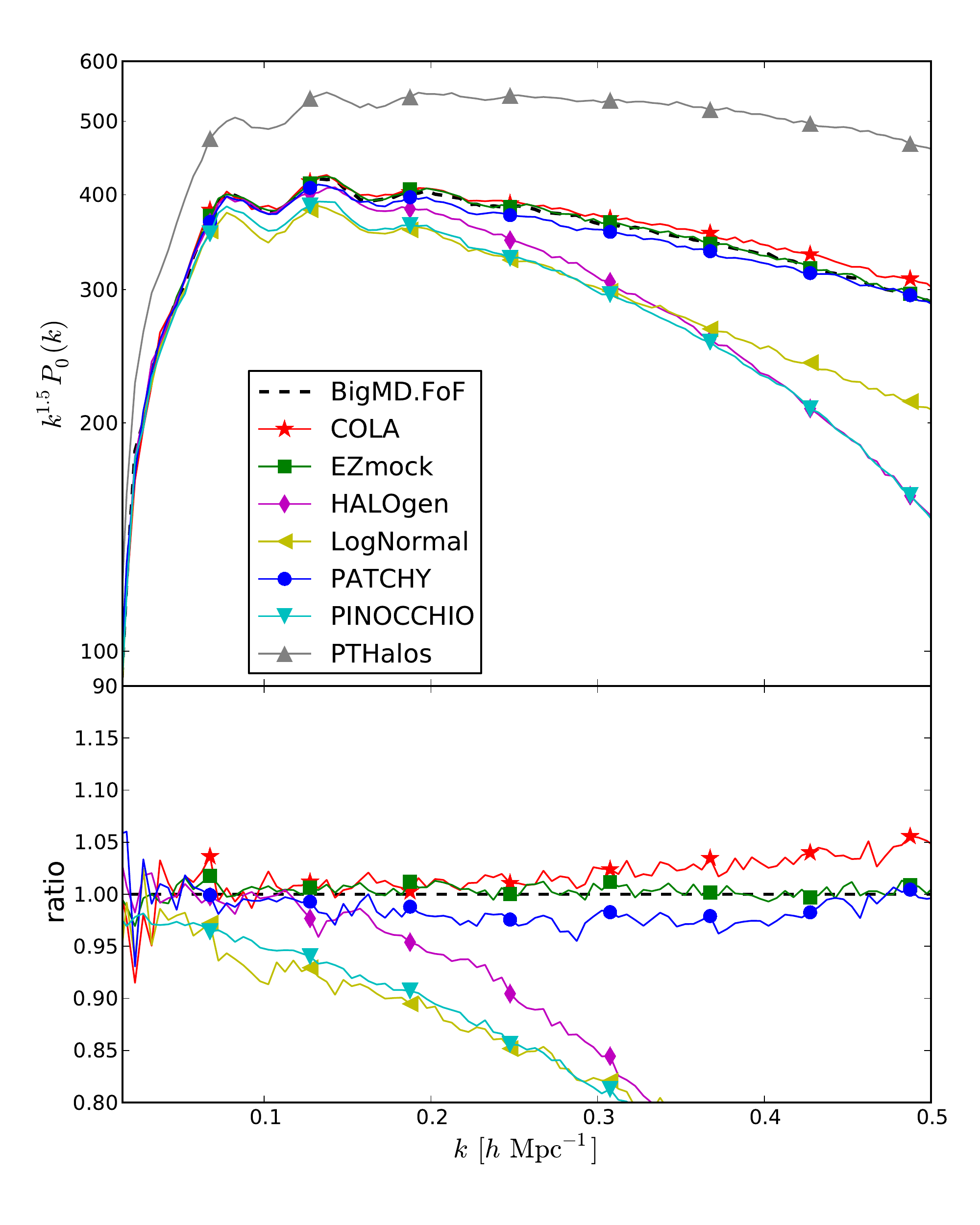}
\caption{FoF power spectrum comparison, in real space, between the different approximate  methods and BigMultiDark.}
\label{fig:FOF_pk_r}
\end{figure}

\begin{figure}
\begin{center}
 \subfigure{\label{fig:pk_z_mono}\includegraphics[width=0.50 \textwidth]{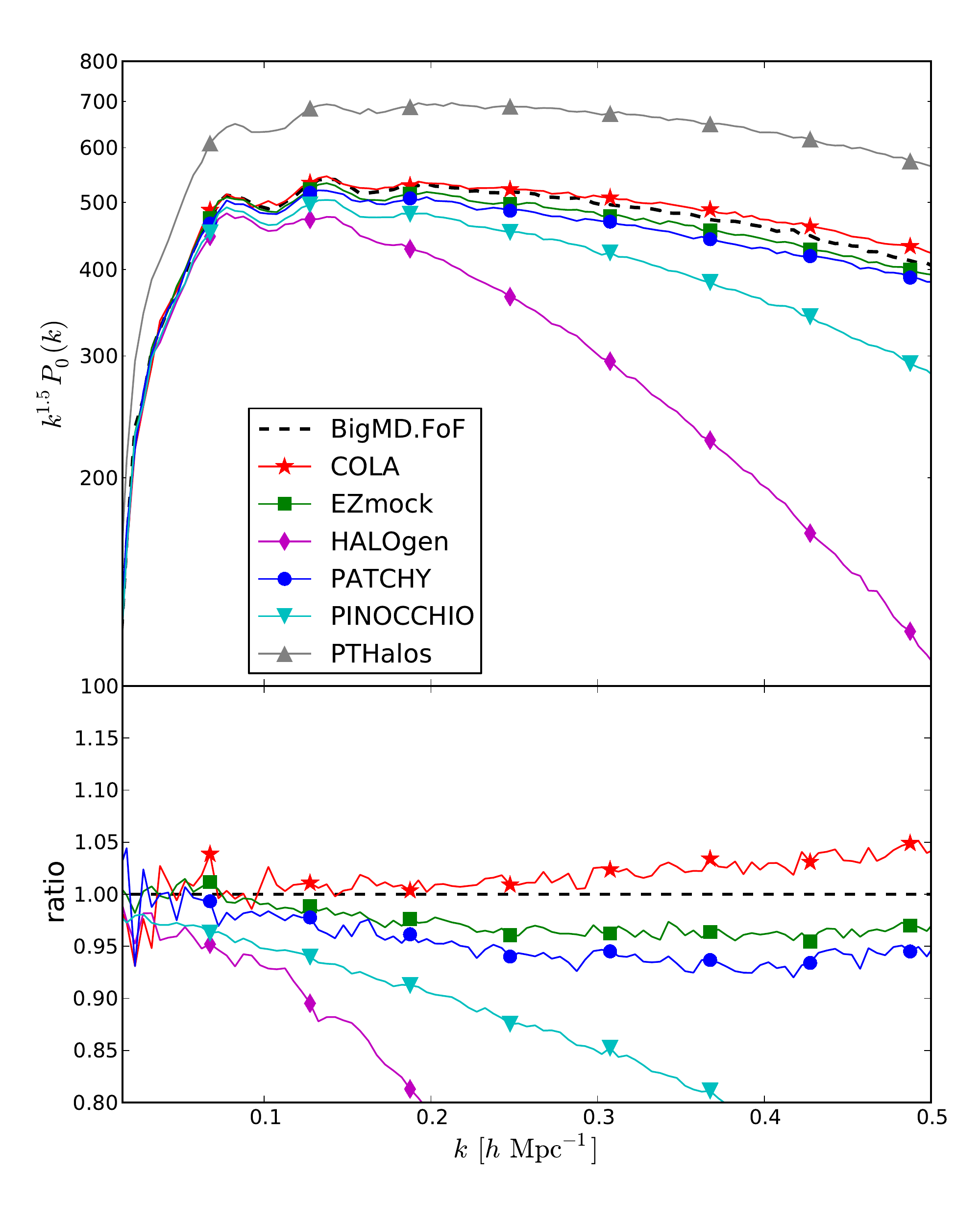}}
 \subfigure{\label{fig:pk_z_quad}\includegraphics[width=0.50 \textwidth]{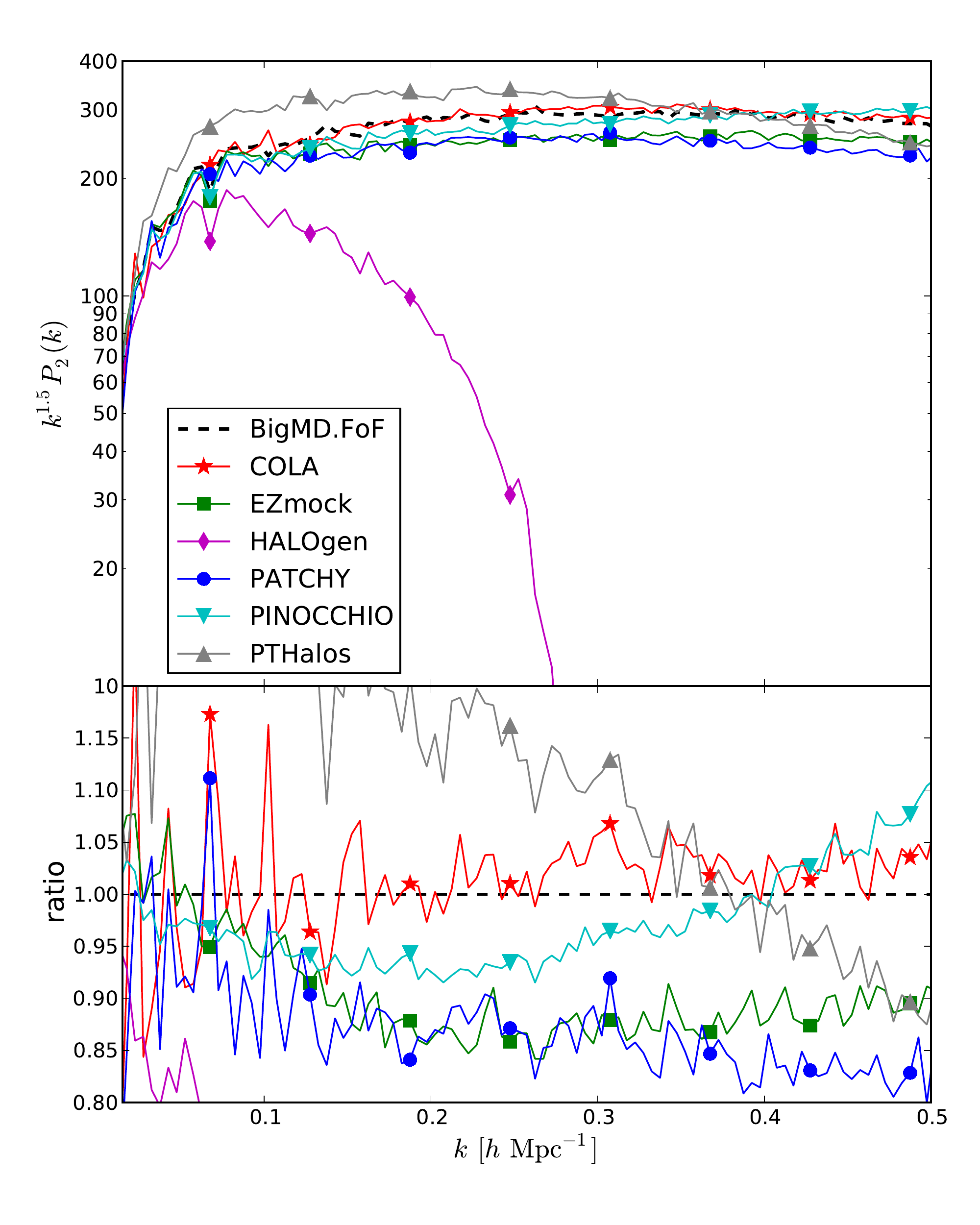}}
\end{center}
\caption{Top panel: performance results for the monopole of the power spectrum in redshift space. Bottom panel: comparison of the quadrupole of the power spectrum in redshift space. Dashed lines correspond to the BigMultiDark FoF reference catalogue.}
\label{fig:FOF_pk_z}
\end{figure}

Although, theoretically, the power spectrum is simply a Fourier transform of the 2-point correlation function, the performance can be very different. The uncertainties at small scales in the configuration space will propagate to the relative large scales in Fourier space.
Fig. \ref{fig:FOF_pk_r} shows the monopole of the power spectrum in real-space. 
COLA, EZmock, and PATCHY agree with the simulation within $5\%$ for all the scales. 
HALOgen, Log-Normal, and PINOCCHIO agree with the simulation within $10\%$ up to $k=0.2-0.25\hompc$.
PTHalos has $\sim20\%$ deviation on the linear bias and we have checked that the deviation of PTHalos can be much smaller if we use lower number density (i.e. massive haloes). In this run the smaller halos have mass equivalent to $\sim10$ particles and some spurious halos are assigned around large overdensities thus increasing the clustering.
Note that the Log-Normal mock is constructed with a input correlation function which is adjusted to be close to that from the simulation. The power spectrum should be better restored if one use a proper input power spectrum.
Fig. \ref{fig:FOF_pk_z} shows the monopole and quadrupole of the power spectrum in redshift-space. 
For the monopole, COLA, EZmock, and PATCHY agree with the simulation within $5\%$ for all the scales shown in the plot;
for the quadrupole, COLA agrees with the simulation within $5\%$ for all the scales; PINOCCHIO agrees within $10\%$; EZmock and PATCHY agree with the simulation within $15-20\%$
We find that only the semi-$N$-body simulation, i.e. COLA, could reach high accuracy at small scales, i.e., $r<25\mpcoh$ or $k>0.15\hompc$, on the quadrupole of the correlation function or the power spectrum. The methods based on perturbation theory seem to have some difficulty improving the precision of quadrupole at small scales.

\subsubsection{3-point clustering statistics of FoF catalogues}
\begin{figure}
\begin{center}
 \subfigure{\label{fig:FOF_3pcf}\includegraphics[width=0.50 \textwidth]{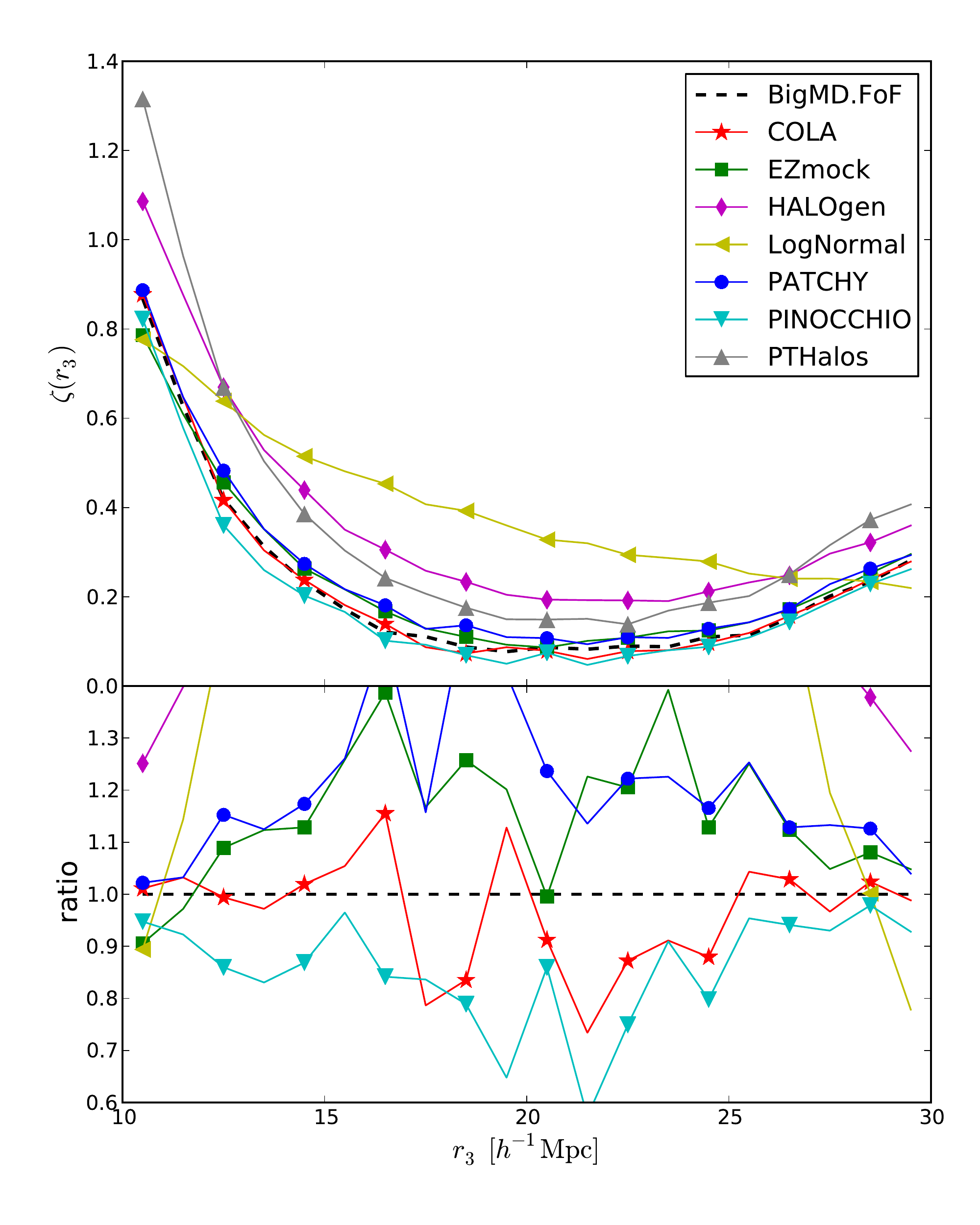}}
 \subfigure{\label{fig:FOF_bik}\includegraphics[width=0.50 \textwidth]{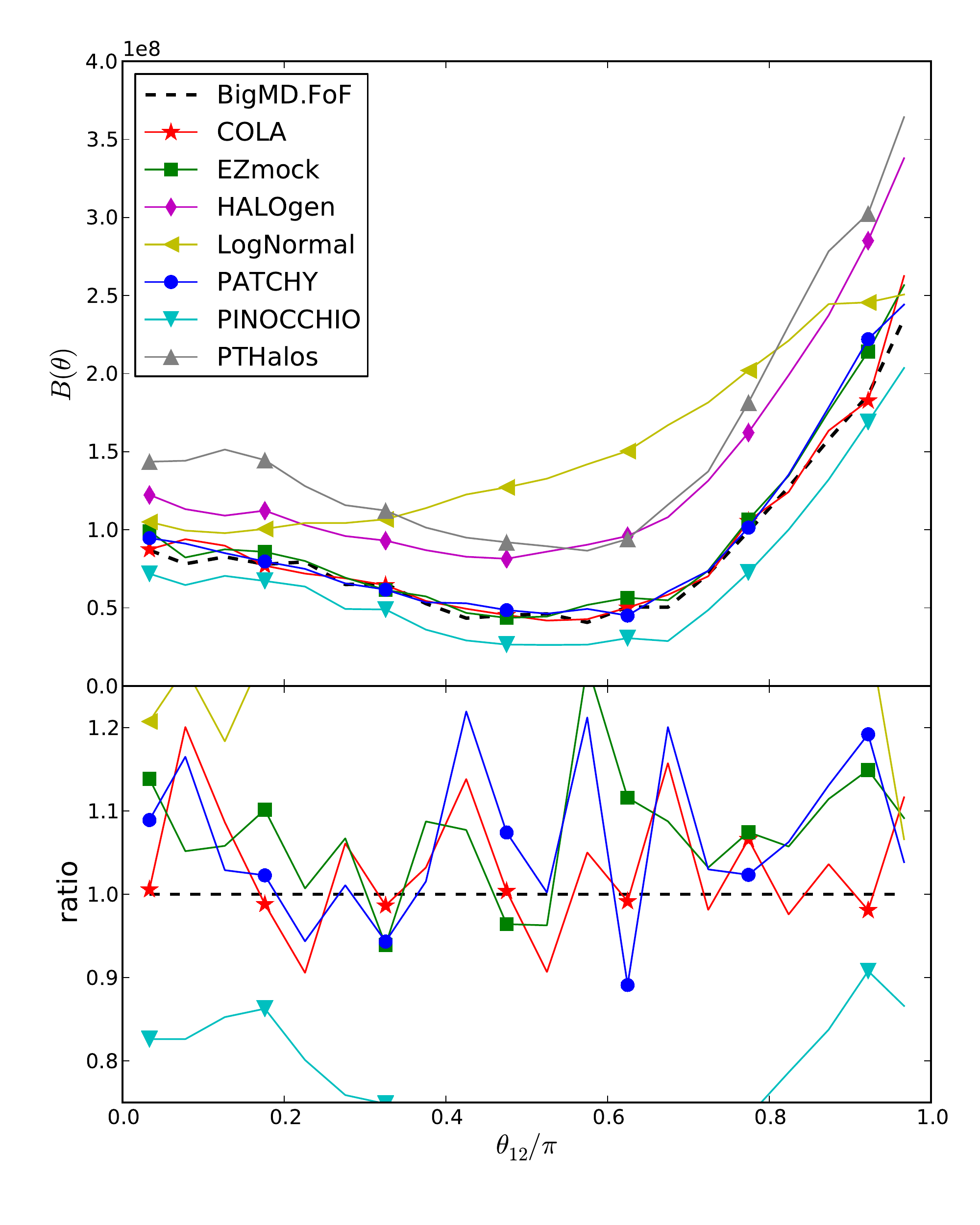}}
\end{center}
\caption{Top panel: performance results for the 3-point correlation function in real-space.. Bottom panel: bispectrum in real-space. Dashed lines correspond to the BigMultiDark FoF reference catalogue.}
\label{fig:FOF_3pt}
\end{figure}

Fig. \ref{fig:FOF_3pt} shows the bispectrum and 3-point correlation function in real space.
To compute 3PCF, we use the ntropy-npoint software, an exact n-point
calculator which uses a kd-tree framework with true parallel
capability and enhanced routine performance \citep{Gardner:2007ua, McBride:2010zn}.
We compute the 3-point correlation functions with the configuration of the triangles with two fixed sides, $r_1=10\mpcoh$ and $r_2=20\mpcoh$, and varying the third side, $r_3$. COLA, EZmock, PATCHY, PINOCCHIO, and PTHalos agree with the simulation within the level of noise.
We compute the bispectrum with the configuration of the triangles given two fixed sides, $k_1=0.1\hompc$ and $k_2=0.2\hompc$, and a varying angle $\theta_{12}$.
COLA, EZmock, and PATCHY agree very well with the reference simulation catalogue. 
We conclude that an appropriate bias model is the key to reach high accuracy for the power spectrum and 3-point clustering statistics.

\subsection{Mocks for SO/BDM catalogues}
Here, we discuss the performance of the different approximate methods when we compare with the spherical overdensity catalogue 
(obtained using BDM code) 
from BigMultiDark with the same halo number density. Note that this catalogue includes both distinct haloes and subhaloes (see Section \ref{sec:bigmd}).
The mesh size used for computing the statistics is $960^3$ if applicable. 
Note that while EZmock, Log-Normal, and PATCHY mocks for the SO catalogue are generated with the same procedures as that for the FoF catalogue, COLA, HALOgen, and PINOCCHIO are including subhaloes following a Halo Occupation Distribution (HOD) scheme described in the Appendix. In addition, while COLA and PINOCCHIO are using the FoF mocks as the distinct haloes to assign the subhaloes arround them, HALOgen constructs a new catalogue matching the SO distinct haloes before the HOD process. PTHalos is not included in this section.

\subsubsection{2-point clustering statistics of SO catalogues}
\begin{figure}
\centering
\includegraphics[width=0.5\textwidth]{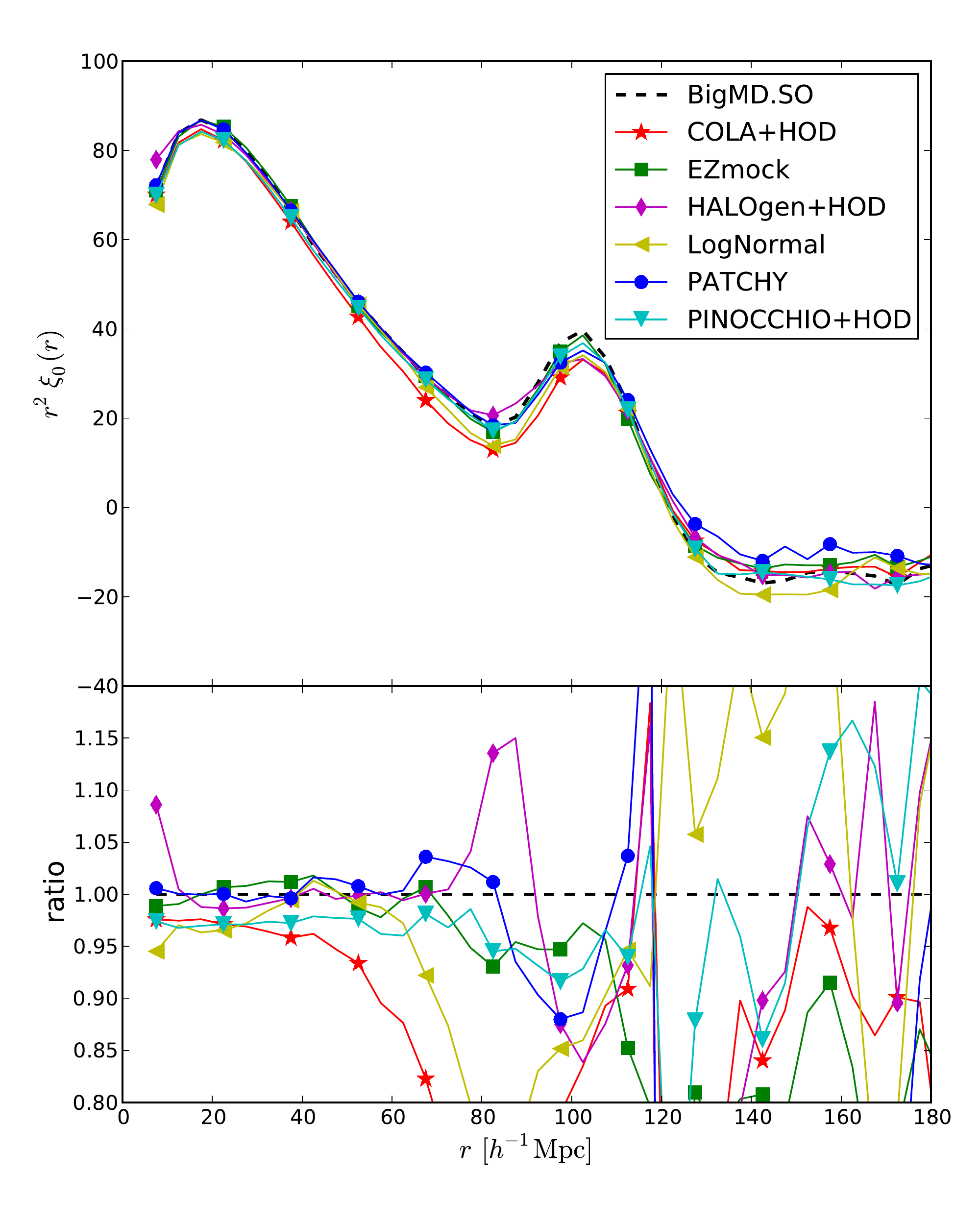}
\caption{Comparison of the monopole of the correlation function in real-space. Dashed line corresponds to the BigMultiDark SO reference catalogue.}
\label{fig:BDM_cf_r}
\end{figure}

\begin{figure}
\begin{center}
 \subfigure{\label{fig:cf_z_mono}\includegraphics[width=0.50 \textwidth]{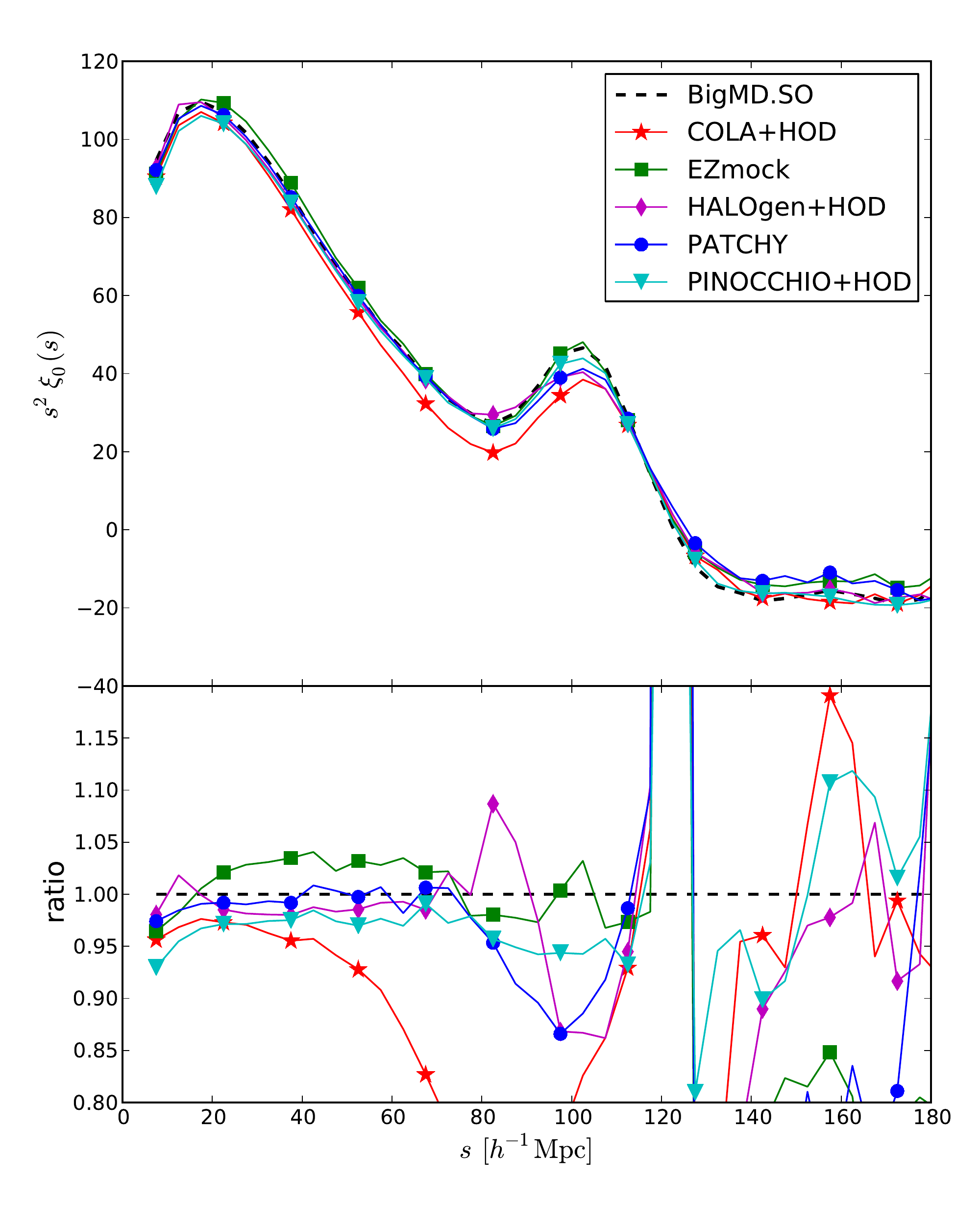}}
 \subfigure{\label{fig:cf_z_quad}\includegraphics[width=0.50 \textwidth]{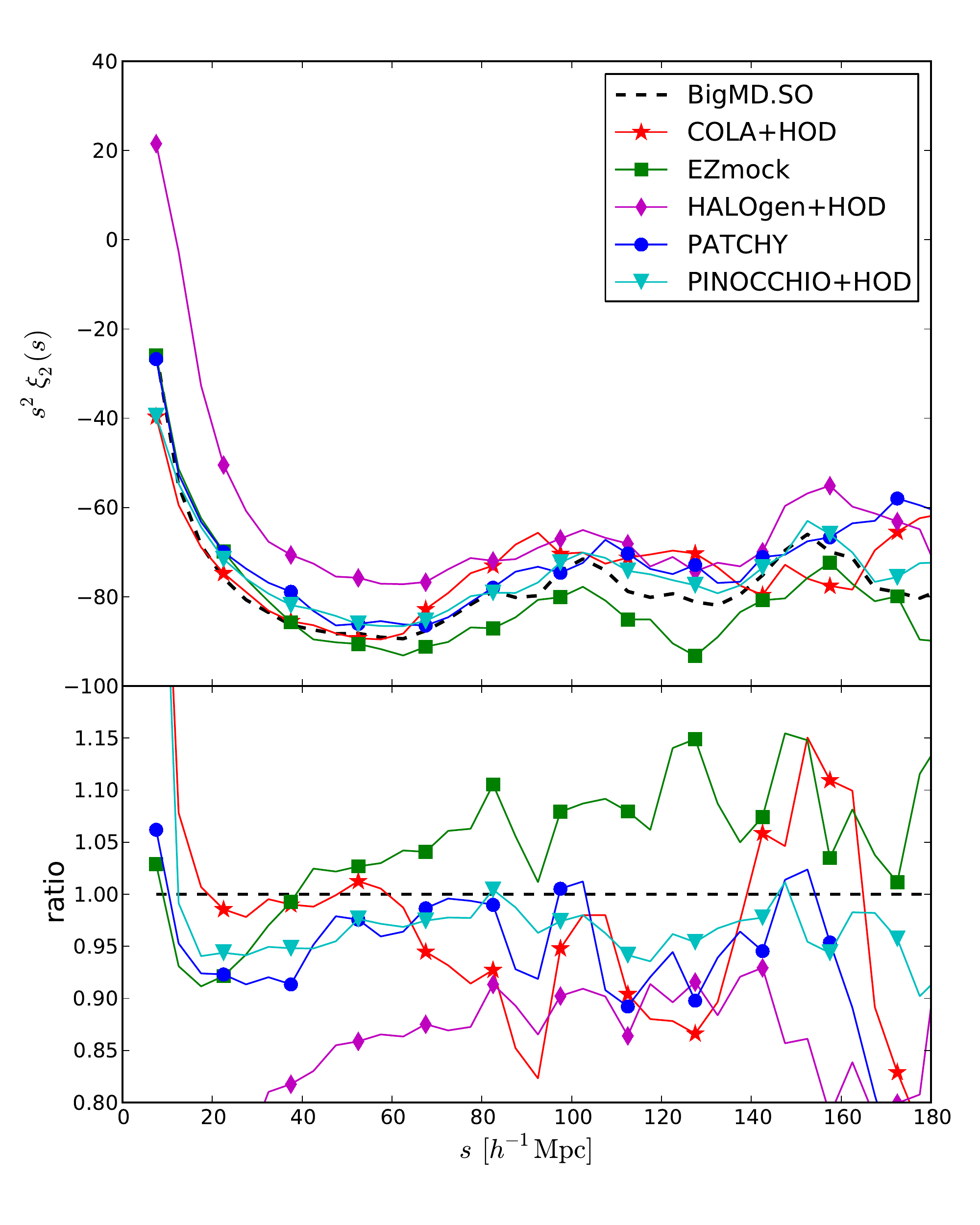}}
\end{center}
\caption{Top panel: comparison of the monopole of the correlation function in redshift space. Bottom panel: performance results for the quadrupole of the correlation function in redshift-space. Dashed lines correspond to the BigMultiDark SO reference catalogue.}
\label{fig:BDM_cf_z}
\end{figure}

Fig. \ref{fig:BDM_cf_r} shows the performance of the different methods on the monopole of correlation function in real-space.
All the mocks agree with the simulation very well.
Fig. \ref{fig:BDM_cf_z} shows the comparison for the monopole and quadrupole of the correlation function in redshift-space. For the monopole, COLA+HOD shows some deviation at scales $>40\mpcoh$, which may be due to not using the BigMultiDark white noise. For the quadrupole, EZmock, PATCHY and PINOCCHIO+HOD agree with the simulation catalogue within $10\%$ for all the scales considered. COLA+HOD agrees within $10\%$ down to $r=15\mpcoh$.

\begin{figure}
\centering
\includegraphics[width=0.5\textwidth]{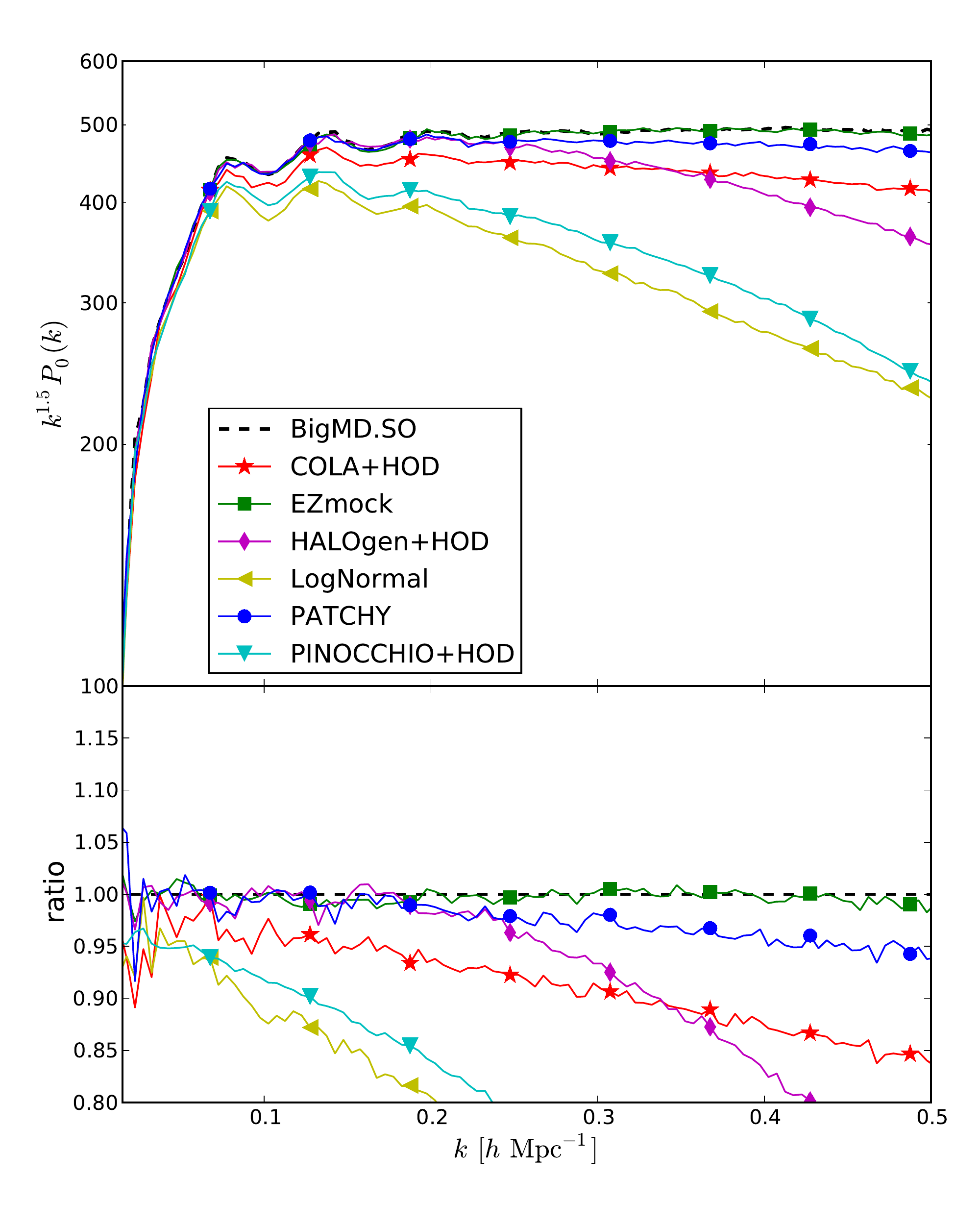}
\caption{SO power spectrum comparison, in real space, between the different approximate  methods and BigMultiDark.}
\label{fig:BDM_pk_r}
\end{figure}

\begin{figure}
\begin{center}
 \subfigure{\label{fig:pk_z_mono}\includegraphics[width=0.50 \textwidth]{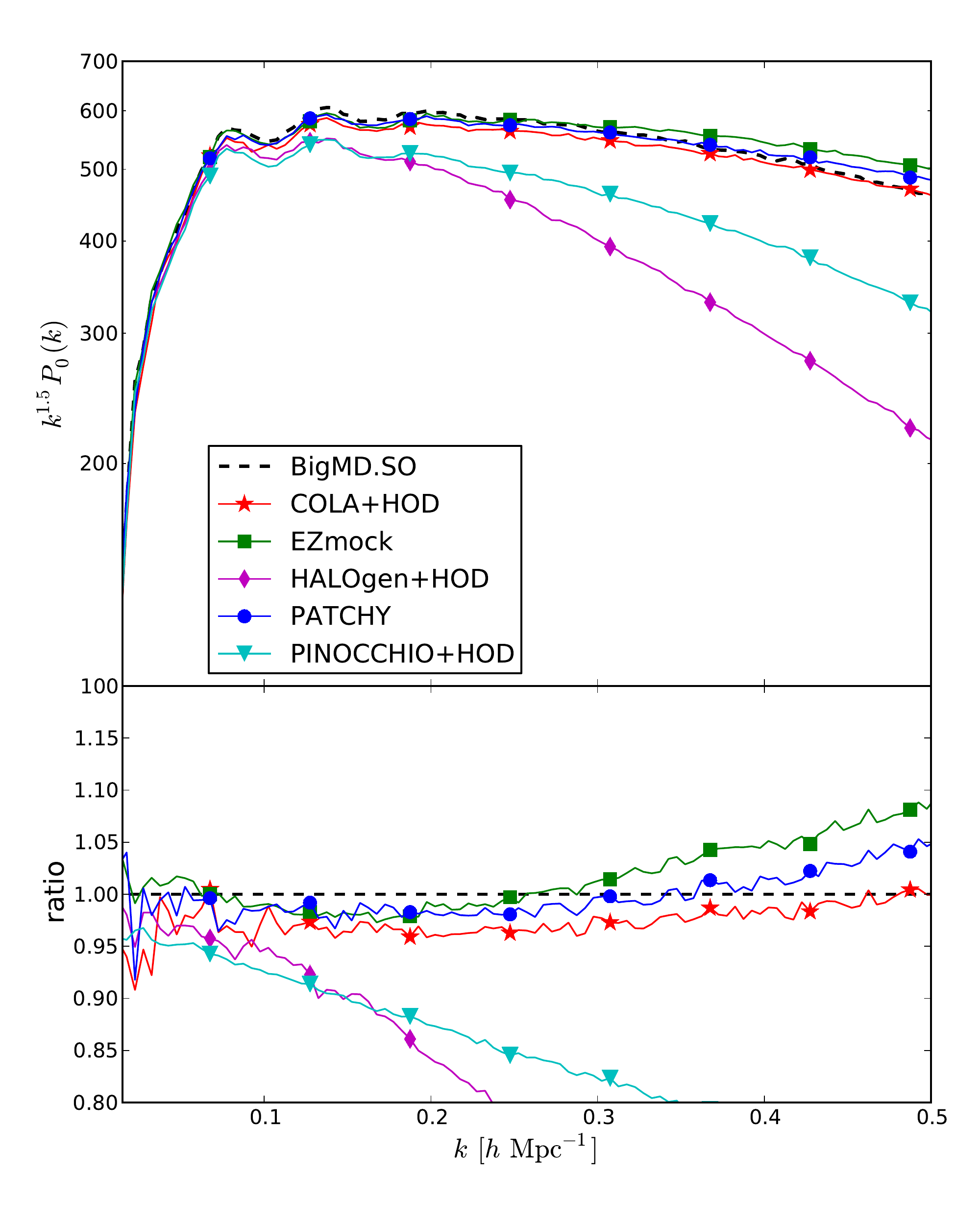}}
 \subfigure{\label{fig:pk_z_quad}\includegraphics[width=0.50 \textwidth]{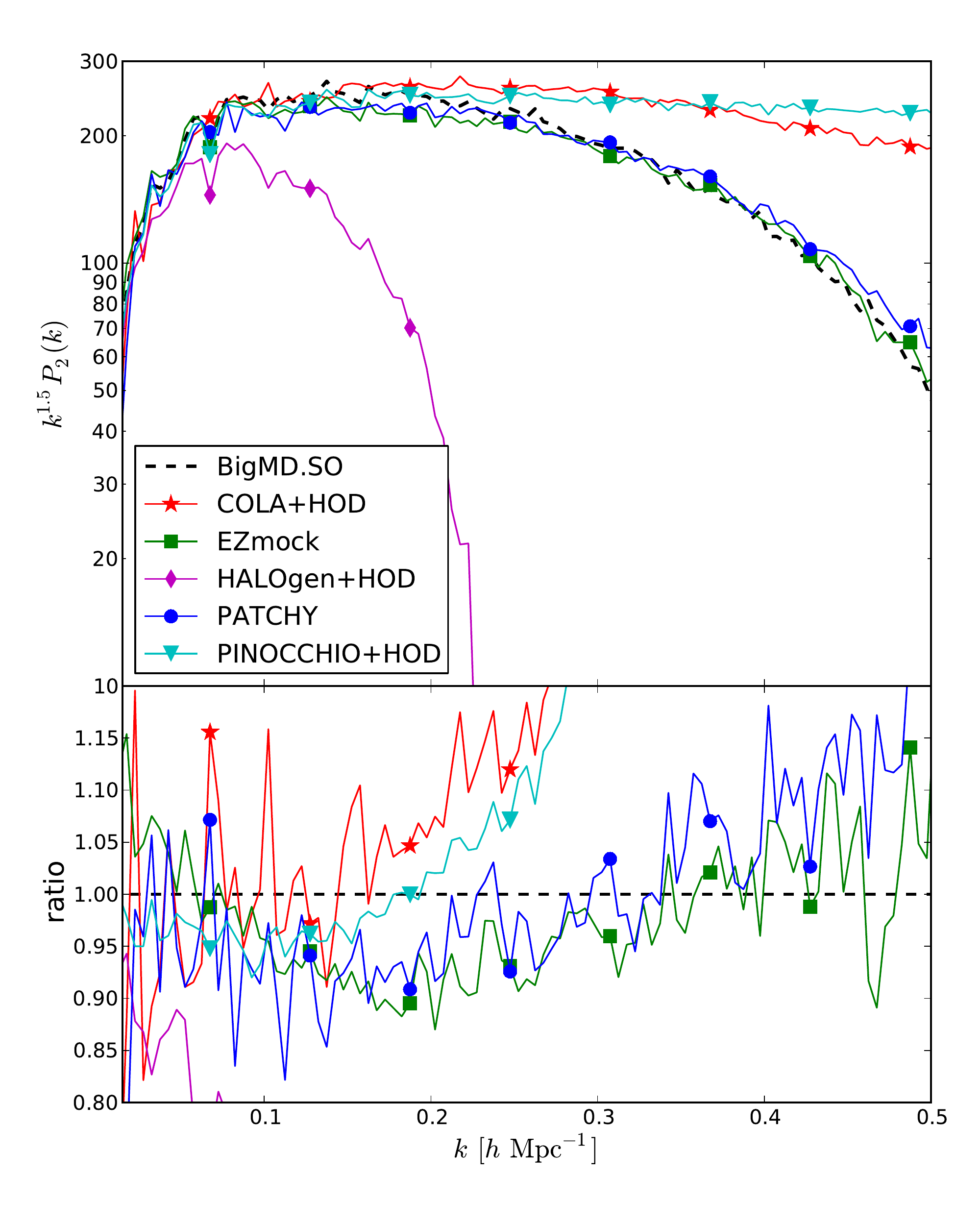}}
\end{center}
\caption{Top panel: performance results for the monopole of the power spectrum in redshift space. Bottom panel: comparison of the quadrupole of the power spectrum in redshift space. Dashed lines correspond to the BigMultiDark SO reference catalogue.}
\label{fig:BDM_pk_z}
\end{figure}

Fig. \ref{fig:BDM_pk_r} shows the monopole of the power spectrum in real-space. EZmock and PATCHY agree with BigMultiDark within $5\%$ for all the scales. COLA+HOD and HALOgen+HOD are within $10\%$ up to $k\sim0.35\hompc$, and PINOCCHIO+HOD and Log-Normal are within $10\%$ up to $k\sim0.1\hompc$. Note again that the Log-Normal mock should be able to agree better with the simulation if one uses a proper input power spectrum.
Fig. \ref{fig:BDM_pk_z} shows the performance comparison for the monopole and quadrupole of the power spectrum in redshift-space. 
COLA+HOD, EZmock, and PATCHY agree with BigMultiDark monopole within $10\%$ for all the scales; ; and up to $k\sim0.1\hompc$ for HALOgen+HOD and PINOCCHIO+HOD. For the quadrupole, EZmock and PATCHY agree with the simulation within $10\%$ for all the scales; COLA+HOD and PINOCCHIO+HOD agree up to $k=0.25\hompc$.

As discussed in the Appendix, we test our HOD scheme by applying it to the  SO distinct halos from the BigMD simulation, trying to reproduce the clustering of substructures. We also test on the BigMD FoF catalogue. We find that HOD scheme has good performance in real space but the difference between SO distinct halo catalogue and FoF catalogue would introduce some bias. We also find that it is not trivial to correctly model the velocity distribution of the substructure which results the relatively poor performance of the HOD model in redshift space.

\subsubsection{3-point clustering statistics of SO catalogues}
\begin{figure}
\begin{center}
 \subfigure{\label{fig:BDM_3pcf}\includegraphics[width=0.50 \textwidth]{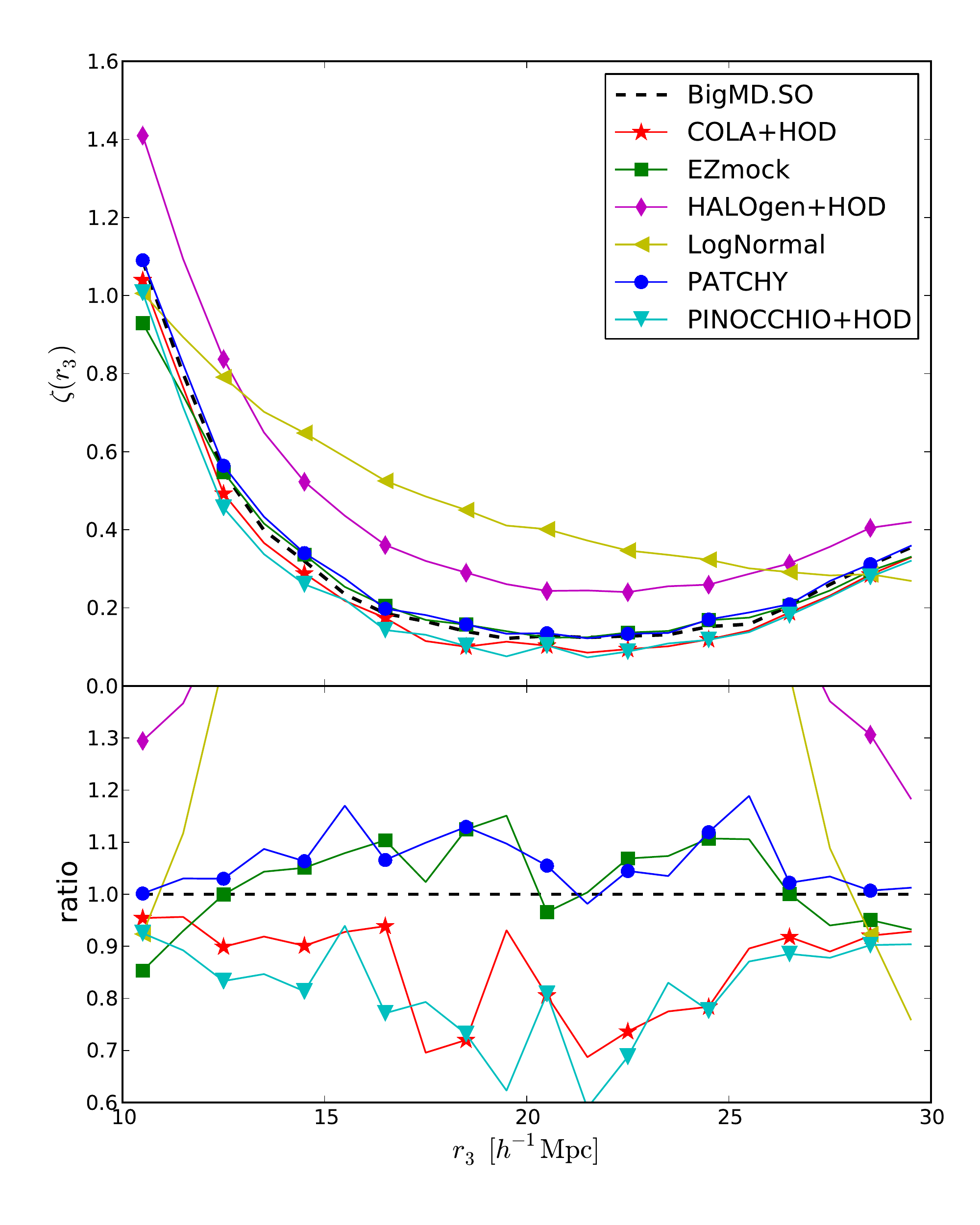}}
 \subfigure{\label{fig:BDM_bik}\includegraphics[width=0.50 \textwidth]{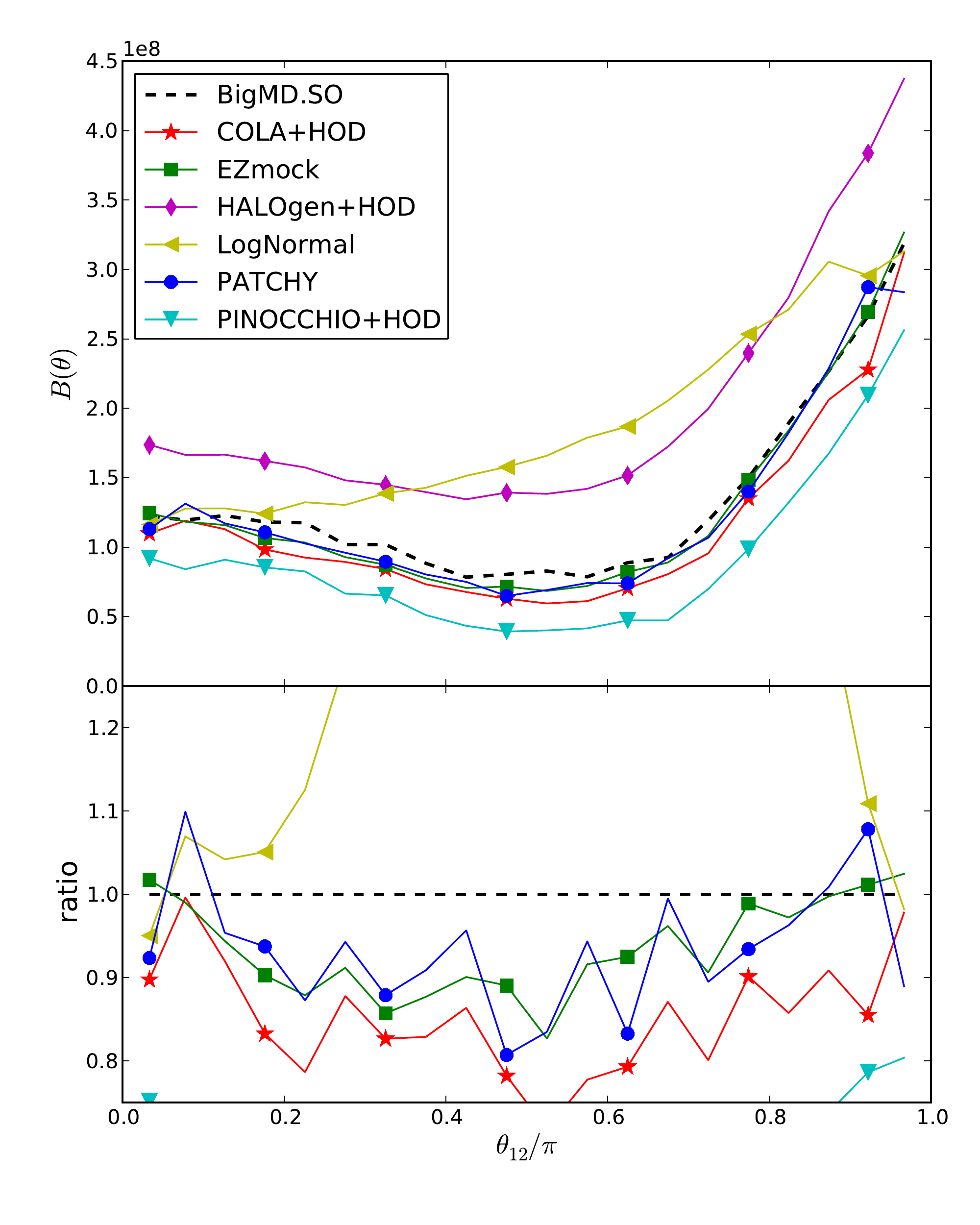}}
\end{center}
\caption{Top panel: performance results for the 3-point correlation function in real-space. Bottom panel: bispectrum in real-space. Dashed lines correspond to the BigMultiDark SO reference catalogue.}
\label{fig:BDM_3pt}
\end{figure}

Fig. \ref{fig:BDM_3pt} shows the bispectrum and 3-point correlation function in real space.
The configurations are the same as for FoF catalogues. 
For the 3-point correlation function, EZmock and PATCHY agree with the simulation within the level of noise. COLA+HOD and PINOCCHIO+HOD agree with the simulation within 20\%.
For the bispectrum, COLA+HOD, EZmock, and PATCHY agree within 10-20\% with the reference simulation catalogue. 
We conclude that an appropriate bias model is the key to reach high accuracy for the power spectrum and 3-point clustering statistics.

\section{SUMMARY}\label{sec:conclusion}
In this paper we have compared the performance of seven different approximate methods to model the halo/galaxy clustering statistics. The resulting mock catalogues from each method have been compared to a reference FoF and SO halo catalog drawn from the Planck BigMultiDark simulation with similar clustering properties that the BOSS galaxies at $z\sim0.5$. Note that the methods compared in this study might have different advantages and applications, e.g., merging history, etc., which are not included in this study.  

We are listing some items we have learned from this comparison study and have more discussion following the list.
\begin{enumerate}
\item[(I)] Most of the methods are able to reproduce the 2-point statistics in configuration space but not necessary in Fourier space,
\item[(II)] an appropriate bias model is the key to reach high accuracy for the power spectrum and 3-point statistics, including bispectrum and 3-point correlation function,
\item[(III)] in redshift space, so far, only the semi-$N$-body simulation, i.e. \textsc{COLA}, could reach high accuracy ($1\%$ level) at small scales, i.e., $r<25\mpcoh$ or $k>0.15\hompc$, on the quadrupole of the correlation function or the power spectrum. 
\item[(IV)] it is not trivial to fit a catalogue that contains substructures (e.g. SO catalogue) starting from a catalogue with only distinct haloes and applying a HOD scheme on it.
\end{enumerate}

The position of dark matter particles after cosmic evolution according to perturbative approaches show a typical uncertainty of roughly a few Mpc,  depending on the chosen approximation (e.g., see \citealt{Monaco:2013qta,Kitaura:2012tj}). This does not show up so clearly in the correlation function in configuration space, where the small scales are kept separated from the large ones. However, it does have a very clear impact in the power spectrum, as it does not reproduce the one halo-term, and thus lacks the commonly known non-linear bump towards high $k$s. Small scale uncertainties propagate in Fourier space having the effect of a convolution (see \citealt{Tassev:2011ac,Monaco:2013qta}). In this work we have presented two kinds of approaches based on perturbation theory. Those which rely on the approximate position of the dark matter particles to find the haloes, and those which just use its large scale structure density field combined with a statistical population prescription to populate the haloes. We find that the first ones are more sensitive to the uncertainty in the particle positions and thus show a larger deviation in Fourier space than in configuration space. While the second class of methods circumvent the problem, by compensating the deviation with the adopted bias description.
It is arguable whether one wants to maintain the analytical models as they are and accept their uncertainties while having a clear understanding of their systematics; or modify them with additional prescriptions to fit the simulations, and introduce more complex relations.

The methods based on perturbation theory seem to have some difficulty improving the precision of quadrupole at small scales. \cite{White:2014gfa} built the theoretical model for biased tracers (i.e. haloes or galaxies) in configuration space and also found similar deviations in the quadrupole comparing to the $N$-body simulation at small scales. 

A HOD model is typically used to analyze some 
two-point clustering measurement (e.g., projected correlation function) and therefore the model is consistent 
with the clustering by construction.  However, one could simply adopt an HOD 
model from a particular halo catalog, and there is no guarantee that the 
resulting mock catalog reproduces the expected clustering signal.
In addition, if a model is calibrated only to 
the clustering length or bias (i.e., the 2-halo term), it might not reproduce the 
small-scale clustering.  Also, different types of galaxies (or haloes) may have 
different spatial clustering and may occupy haloes differently or have different 
central/satellite fractions, so it's important to note that different HOD models 
may be required.
While our HOD application leads to the results reported in this study, an improved (less standard or less straightforward) application could yield better agreement in terms of two-point statistics. This should further investigated in future works.

\section{Acknowledgement}
The authors would like to express special thanks to the Instituto de Fisica Teorica (IFT-UAM/CSIC in Madrid) for its hospitality and support, via the Centro de Excelencia Severo Ochoa Program under Grant No. SEV-2012-0249, during the three week workshop `nIFTy Cosmology' where this work developed. We further acknowledge the financial support of the University of Western 2014 Australia Research Collaboration Award for `Fast Approximate Synthetic Universes for the SKA', the ARC Centre of Excellence for All Sky Astrophysics (CAASTRO) grant number CE110001020, and the two ARC Discovery Projects DP130100117 and DP140100198. We also recognize support from the Universidad Autonoma de Madrid (UAM) for the workshop infrastructure.

We especially like to thank Frazer Pearce for initiating (together with AK) the Mocking Astrophysics\footnote{http://www.mockingastrophysics.org} program under whose umbrella the workshop and this work was performed, respectively.
We thank the developers of COLA, Tassev, Zaldarriaga, Eisenstein and Koda, for making the code public available.
We thank Jeffery Gardner and Cameron McBride for sharing the ntropy-npoint code.

CC and FP were supported by the Spanish MICINN’s Consolider-Ingenio 2010 Programme under grant MultiDark CSD2009-00064 and AYA2010-21231-C02-01 grant, and Spanish MINECO’s “Centro de Excelencia Severo Ochoa” Programme under grant SEV-2012-0249.  
CZ and CT acknowledges support from Tsinghua University, and 973 program No. 2013CB834906. CZ also thanks the support from the MultiDark summer student program to visit the Instituto de F\'{\i}sica Te\'orica, (UAM/CSIC), Spain.
AK is supported by the {\it Ministerio de Econom\'ia y Competitividad} (MINECO) in Spain through grant AYA2012-31101, as well as the Consolider-Ingenio 2010 Programme of the {\it Spanish Ministerio de Ciencia e Innovaci\'on} (MICINN) under grant MultiDark CSD2009-00064. He also acknowledges support from the {\it Australian Research Council} (ARC) grants DP130100117 and DP140100198. 
EM and PM have been supported by a FRA2012 grant of the University of Trieste, PRIN2010-2011 (J91J12000450001) from MIUR, and by Consorzio per la Fisica di Trieste.
AI is supported by the JAE program grant from the Spanish National Science Council (CSIC).
GY acknowledges support from the Spanish MINECO under research grants  AYA2012-31101, FPA2012-34694, AYA2010-21231, Consolider Ingenio SyeC CSD2007-0050 and  from Comunidad de Madrid under  ASTROMADRID  project (S2009/ESP-1496).
VT and SG have been supported by the German Science Foundation under DFG grant GO563/23-1.
FAM is supported by the Australian Research Council Centre of Excellence for All-Sky Astrophysics (CAASTRO) through project number CE110001020.

The MultiDark Database (www.multidark.org) used in this paper and the web application providing online access to it were constructed as part of the activities of the German Astrophysical Virtual Observatory as result of a collaboration between the Leibniz-Institute for Astrophysics Potsdam (AIP) and the Spanish MultiDark Consolider Project CSD2009-00064.  The  BigMD simulation suite have been performed in the Supermuc supercomputer at LRZ using time granted by PRACE. The simulation has been performed with an optimized version of GADGET-2 kindly provided by Volker Springel.
We also acknowledge PRACE for awarding us access to resource Curie supercomputer based in France (project PA2259). 
PTHalos were run in COSMA, the HPC facilities at the ICC in Durham, for a project granted by of the STFC DiRAC HPC Facility (www.dirac.ac.uk).
Some other computation were performed on HYDRA, the HPC-cluster of the IFT-UAM/CSIC.

The authors contributed in the following ways to this paper: 
FP coordinated the approximate method workshop programme from which this comparison project and paper originated. 
The analysis presented here was performed by CC and CZ. The paper was written by CC, EM, and FP. 
The mock catalogues and descriptions are run and provided by AI (COLA), SA (HALOgen), CC (EZmock), FK (PATCHY), MM (PTHalos), PM (PINOCCHIO), and SM (HALOgen).
AK, AAK, CS, GY SG, and VT prepared the reference BigMultiDark simulation catalogues;
EM developed and applied a HOD scheme for this study;
VM and FM helped to develop the code for computing 3-point correlation function; 
other authors contributed towards the content of the paper and helped to proof-read it.

\appendix
\section{Assigning subhalos with a HOD prescription}
The approximative mock methods are all designed to give halo catalogues, but (due to aforementioned limitations) not all of them are capable of adding subhaloes to them. Therefore we applied a post-processing step, i.e. the halo occupation distribution, to them augmenting their submitted catalogues with subhaloes. 
The Halo Occupation Distribution (HOD) approach is based on a statistical assignment of the number,
positions, and velocities of substructures residing in a halo as a
function of the halo mass, e.g., \cite{Berlind:2001xk, Kravtsov:2003sg, Zheng:2004id, Skibba:2008gr, Zehavi:2010bh}.

We have applied an HOD scheme to PINOCCHIO, COLA, and \textsc{HALOgen} halo
catalogues. For the first two methods, we have first converted the
values of mass into the values corresponding to bound masses, in order
to be compatible with the definition adopted in the BigMultiDark
simulation. For PINOCCHIO and COLA, we have looked for a
transformation that maps the halo masses into new mass values imposing
that the mass function matches the one of the BigMD SO reference catalogue.

The following step consists in looking for a relation that associates
the halo mass of the BigMultiDark with the average number of
substructures in the halos of that mass. 

We have considered logarithmically equispaced mass bins. In each bin
the distribution of halos with a given number of substructures (main
halos included) is verified to be Poisson distributed, and the best
fit Poisson parameter $\lambda(M)$ is assigned to that bin as
representative of the mean number of substructures.

It is now possible to populate the halos obtained with PINOCCHIO, COLA
and HALOgen, with a population of substructures statistically
identical to that of the BigMultiDark reference catalogue. The actual
number of substructures in a halo is assigned as a random number taken
from a Poisson distribution having the mean value $\lambda(M)$.

Substructures are spatially distributed in order to have an NFW number
density profile, with concentration equal to the main halo's one. The
latter is computed following Bhattacharya, et al. 2013. Peculiar
velocities in each of the three directions are randomly extracted from
a Gaussian distribution having null mean and dispersion equal to
$\sqrt{G M(r)/r}$.

\begin{figure}
\centering
\includegraphics[width=0.5\textwidth]{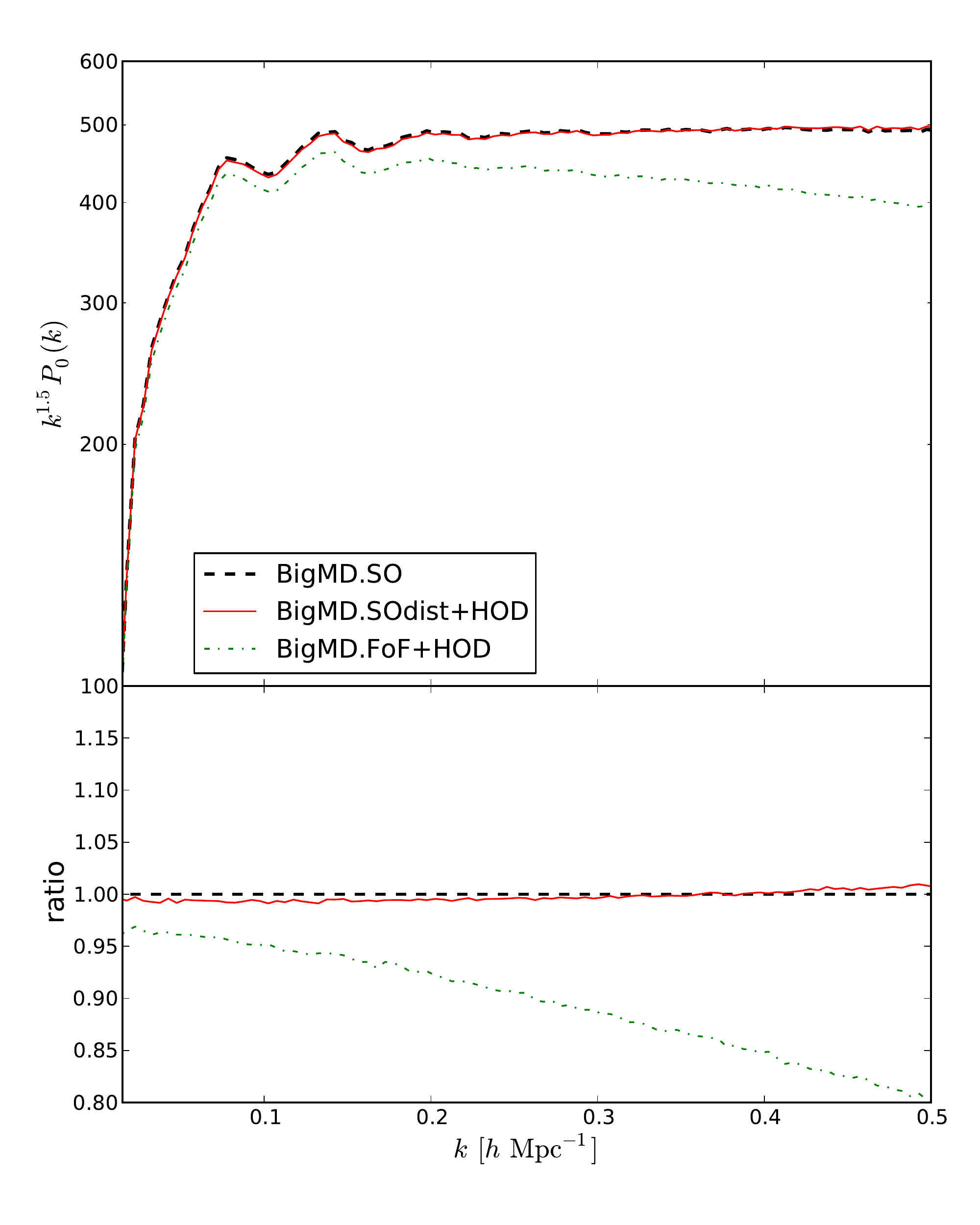}
\caption{HOD Power spectrum comparison, in real space, among the BigMultiDark SO catalogue, SO distinct halos catalogue with HOD applied, and FoF catalogue with HOD applied.}
\label{fig:HOD_pk_r}
\end{figure}

\begin{figure}
\begin{center}
 \subfigure{\includegraphics[width=0.50 \textwidth]{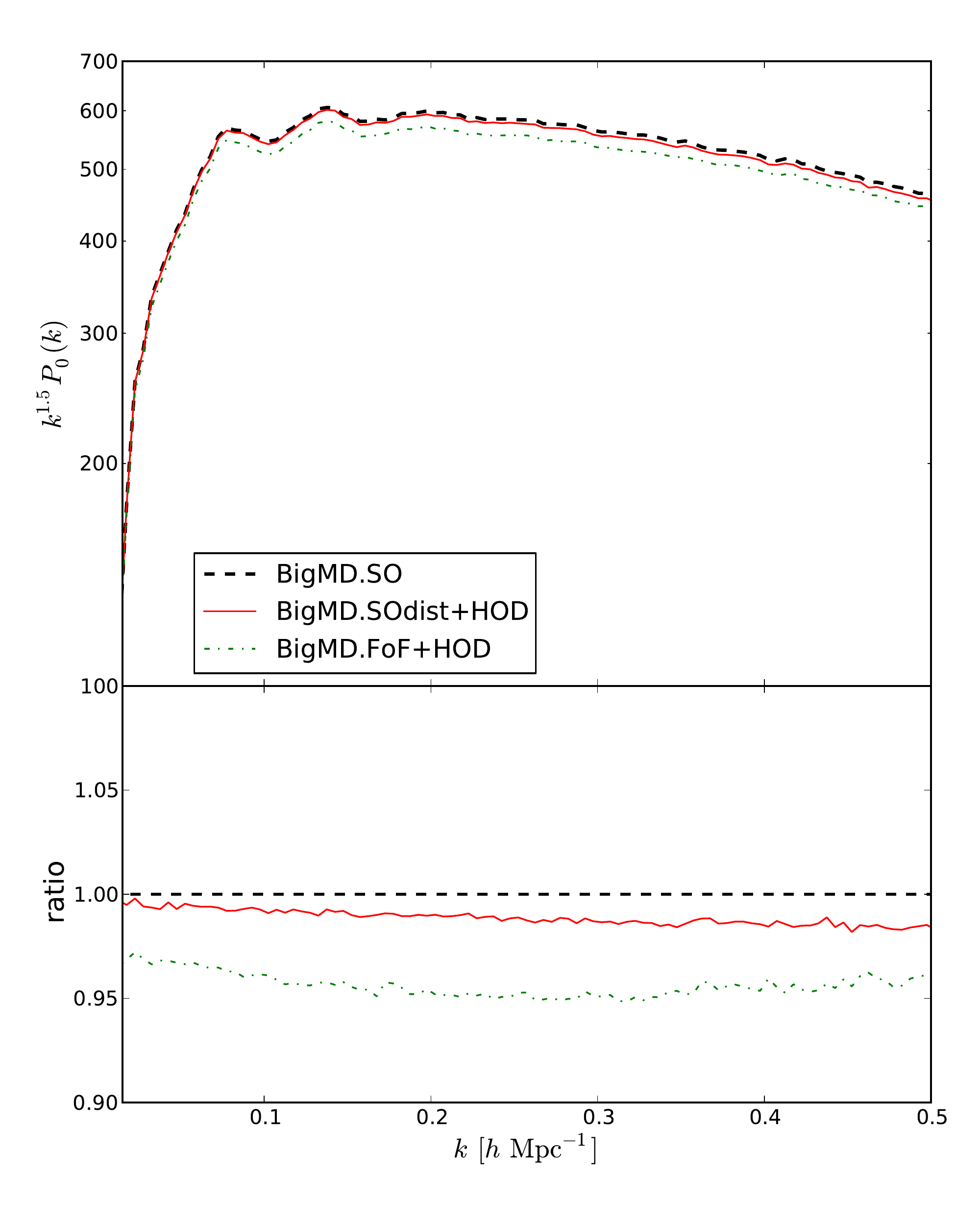}}
 \subfigure{\includegraphics[width=0.50 \textwidth]{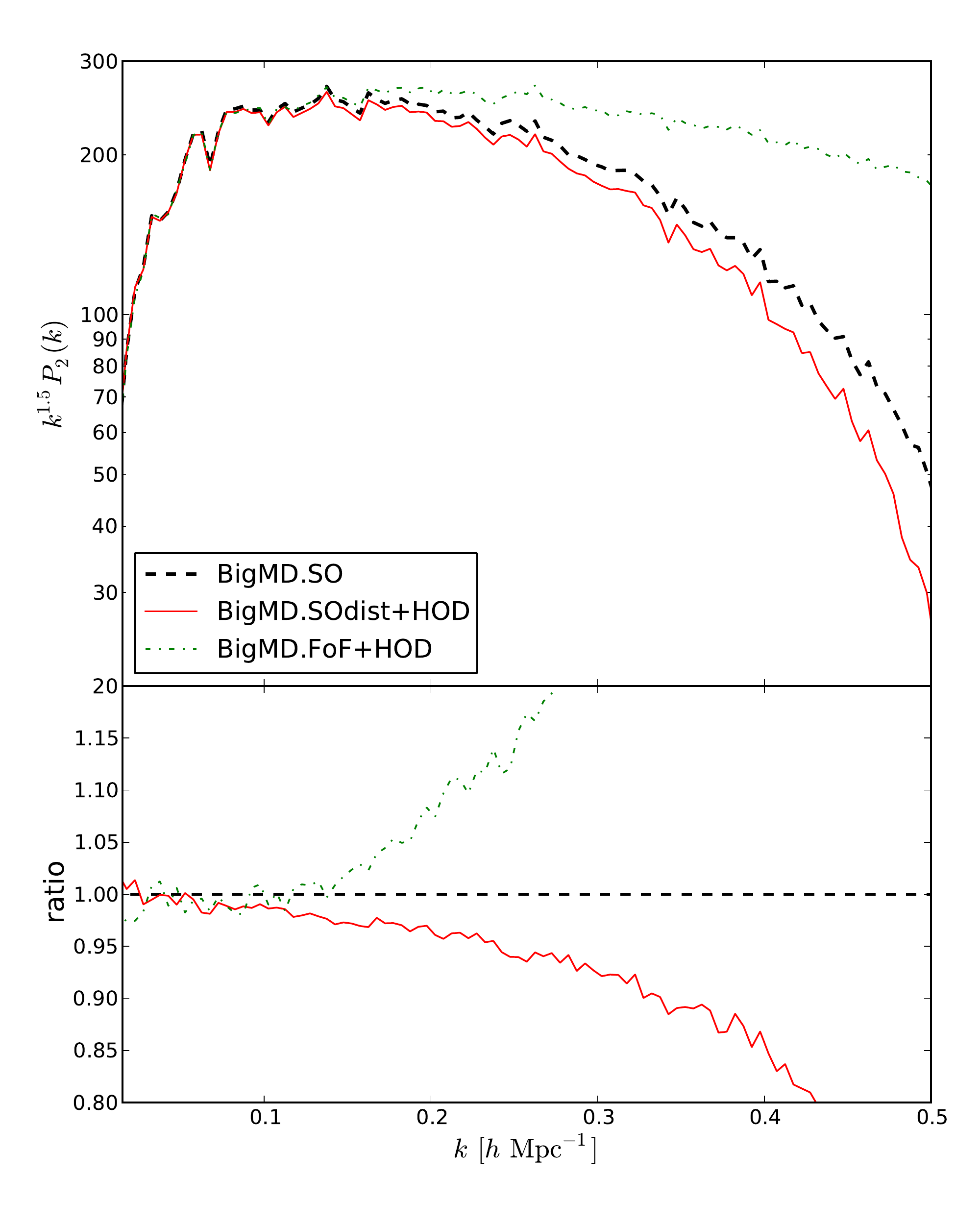}}
\end{center}
\caption{Top panel: HOD performance results for the monopole of the power spectrum in redshift space. Bottom panel: comparison of the quadrupole of the power spectrum in redshift space. 
}
\label{fig:HOD_pk_z}
\end{figure}

We test and validate our HOD scheme by applying it on BigMD SO distinct halo catalogue and BigMD FoF catalogue. Fig. \ref{fig:HOD_pk_r} shows the power spectrum in real space. One can see that BigMD SO distinct halos with HOD scheme applied agrees with the full BigMD SO catalogue very well. BigMD FoF catalog with HOD scheme applied has 5\% deviation which will propagate to the mocks to which we apply the HOD scheme in this study. Fig. \ref{fig:HOD_pk_z} shows the monopole and quadrupole of power spectrum in redshift space. For the monopole, BigMD SO distinct halos with HOD scheme applied agrees with the full BigMD SO catalogue very well; for quadrupole, it agree within 20\% up to $k=0.4 \hompc$.

\bibliography{mockcompare}

\begin{thebibliography}{}

\bibitem[\protect\citeauthoryear{Abazajian et~al.,}{Abazajian
  et~al.}{2009}]{Abazajian:2008wr}
Abazajian K.~N.,  et~al., 2009, ApJS, 182, 543, \eprint{arXiv:0812.0649}

\bibitem[\protect\citeauthoryear{Abell et~al.,}{Abell
  et~al.}{2009}]{Abell:2009aa}
Abell P.~A.,  et~al., 2009, \eprint{0912.0201}

\bibitem[\protect\citeauthoryear{Avila, G.Murray, Knebe, Power, Robotham \&
  Garcia-Bellido}{Avila et~al.}{2014}]{Avila:2014aa}
Avila S.,  G.Murray S.,  Knebe A.,  Power C.,  Robotham A.~S.,
  Garcia-Bellido J.,  2014, \eprint{1412.5228}

\bibitem[\protect\citeauthoryear{Benitez et~al.,}{Benitez
  et~al.}{2014}]{Benitez:2014ibt}
Benitez N.,  et~al., 2014, \eprint{1403.5237}

\bibitem[\protect\citeauthoryear{Berlind \& Weinberg}{Berlind \&
  Weinberg}{2002}]{Berlind:2001xk}
Berlind A.~A.,  Weinberg D.~H.,  2002, ApJ, 575, 587,
  \eprint{astro-ph/0109001}

\bibitem[\protect\citeauthoryear{Beutler, Blake, Colless, Jones, Staveley-Smith
  et~al.,}{Beutler et~al.}{2011}]{Beutler:2011hx}
Beutler F.,  Blake C.,  Colless M.,  Jones D.~H.,  Staveley-Smith L.,
  et~al., 2011, MNRAS, 416, 3017, \eprint{1106.3366}

\bibitem[\protect\citeauthoryear{Blake, Davis, Poole, Parkinson, Brough
  et~al.,}{Blake et~al.}{2011}]{Blake:2011wn}
Blake C.,  Davis T.,  Poole G.,  Parkinson D.,  Brough S.,    et~al., 2011,
  MNRAS, 415, 2892, \eprint{1105.2862}

\bibitem[\protect\citeauthoryear{Bryan \& Norman}{Bryan \&
  Norman}{1998}]{Bryan:1997dn}
Bryan G.,  Norman M.,  1998, ApJ, 495, 80, \eprint{astro-ph/9710107}

\bibitem[\protect\citeauthoryear{Chuang, Kitaura, Prada, Zhao \& Yepes}{Chuang
  et~al.}{2015}]{Chuang:2014vfa}
Chuang C.-H.,  Kitaura F.-S.,  Prada F.,  Zhao C.,    Yepes G.,  2015, MNRAS, 446, 2621, 
  \eprint{1409.1124}

\bibitem[\protect\citeauthoryear{Chuang, Wang \& Hemantha}{Chuang
  et~al.}{2012}]{Chuang:2010dv}
Chuang C.-H.,  Wang Y.,    Hemantha M. D.~P.,  2012, MNRAS,
  423, 1474, \eprint{1008.4822}

\bibitem[\protect\citeauthoryear{Cole et~al.,}{Cole
  et~al.}{2005}]{Cole:2005sx}
Cole S.,  et~al., 2005, MNRAS, 362, 505,
  \eprint{astro-ph/0501174}

\bibitem[\protect\citeauthoryear{Coles \& Jones}{Coles \&
  Jones}{1991}]{Coles:1991if}
Coles P.,  Jones B.,  1991, MNRAS, 248, 1

\bibitem[\protect\citeauthoryear{Colless et~al.,}{Colless
  et~al.}{2001}]{Colless:2001gk}
Colless M.,  et~al., 2001, MNRAS, 328, 1039,
  \eprint{astro-ph/0106498}

\bibitem[\protect\citeauthoryear{Colless, Peterson, Jackson, Peacock, Cole
  et~al.,}{Colless et~al.}{2003}]{Colless:2003wz}
Colless M.,  Peterson B.~A.,  Jackson C.,  Peacock J.~A.,  Cole S.,    et~al.,
  2003, \eprint{astro-ph/0306581}

\bibitem[\protect\citeauthoryear{Davis, Efstathiou, Frenk \& White}{Davis
  et~al.}{1985}]{Davis:1985rj}
Davis M.,  Efstathiou G.,  Frenk C.~S.,    White S.~D.,  1985, ApJ,
  292, 371

\bibitem[\protect\citeauthoryear{Dawson et~al.,}{Dawson
  et~al.}{2013}]{Dawson:2012va}
Dawson K.~S.,  et~al., 2013, AJ, 145, 10, \eprint{1208.0022}

\bibitem[\protect\citeauthoryear{de Jong, Bellido-Tirado, Chiappini, Depagne,
  Haynes et~al.,}{de~Jong et~al.}{2012}]{deJong:2012nj}
de Jong R.~S.,  Bellido-Tirado O.,  Chiappini C.,  Depagne E.,  Haynes R.,
  et~al., 2012, \eprint{1206.6885}

\bibitem[\protect\citeauthoryear{de~la Torre, Guzzo, Peacock, Branchini, Iovino
  et~al.,}{de~la Torre et~al.}{2013}]{delaTorre:2013rpa}
de~la Torre S.,  Guzzo L.,  Peacock J.,  Branchini E.,  Iovino A.,    et~al.,
  2013, A\&A, 557, A54, \eprint{1303.2622}

\bibitem[\protect\citeauthoryear{Drinkwater, Jurek, Blake, Woods, Pimbblet
  et~al.,}{Drinkwater et~al.}{2010}]{Drinkwater:2009sd}
Drinkwater M.~J.,  Jurek R.~J.,  Blake C.,  Woods D.,  Pimbblet K.~A.,
  et~al., 2010, MNRAS, 401, 1429, \eprint{0911.4246}

\bibitem[\protect\citeauthoryear{Eisenstein et~al.,}{Eisenstein
  et~al.}{2011}]{Eisenstein:2011sa}
Eisenstein D.~J.,  et~al., 2011, AJ, 142, 72, \eprint{1101.1529}

\bibitem[\protect\citeauthoryear{Frieman~J.}{Frieman~J.}{2013}]{DES13}
Frieman~J. D. E. S.~C.,  2013, The Dark Energy Survey: Overview.
Vol.~221, American Astronomical Society, AAS Meeting

\bibitem[\protect\citeauthoryear{Gardner, Connolly \& McBride}{Gardner
  et~al.}{2007}]{Gardner:2007ua}
Gardner J.~P.,  Connolly A.,    McBride C.,  2007, \eprint{0709.1967}

\bibitem[\protect\citeauthoryear{Green, Schechter, Baltay, Bean, Bennett
  et~al.,}{Green et~al.}{2012}]{Green:2012mj}
Green J.,  Schechter P.,  Baltay C.,  Bean R.,  Bennett D.,    et~al., 2012,
  \eprint{1208.4012}

\bibitem[\protect\citeauthoryear{Hill, Gebhardt, Komatsu, Drory, MacQueen
  et~al.,}{Hill et~al.}{2008}]{Hill:2008mv}
Hill G.,  Gebhardt K.,  Komatsu E.,  Drory N.,  MacQueen P.,    et~al., 2008,
  ASP Conf.Ser., 399, 115, \eprint{0806.0183}

\bibitem[\protect\citeauthoryear{{Hubble}}{{Hubble}}{1934}]{hubble1934}
{Hubble} E.,  1934, ApJ, 79, 8,
  \adsurl{http://adsabs.harvard.edu/abs/1934ApJ....79....8H}

\bibitem[\protect\citeauthoryear{Kazin, Koda, Blake \& Padmanabhan}{Kazin
  et~al.}{2014}]{Kazin:2014qga}
Kazin E.~A.,  Koda J.,  Blake C.,    Padmanabhan N.,  2014,
  MNRAS, 441, 3524, \eprint{1401.0358}

\bibitem[\protect\citeauthoryear{Kitaura \& Angulo}{Kitaura \&
  Angulo}{2012}]{Kitaura:2011ay}
Kitaura F.,  Angulo R.,  2012, MNRAS, 425, 2443, \eprint{1111.6617}

\bibitem[\protect\citeauthoryear{Kitaura, Gil-Mar{\'\i}n, Scoccola, Chuang,
  M{\"u}ller et~al.,}{Kitaura et~al.}{2014b}]{Kitaura:2014mja}
Kitaura F.-S.,  Gil-Mar{\'\i}n H.,  Scoccola C.,  Chuang C.-H.,  M{\"u}ller V.,
     et~al., 2014, \eprint{1407.1236}

\bibitem[\protect\citeauthoryear{Kitaura \& Hess}{Kitaura \&
  Hess}{2013}]{Kitaura:2012tj}
Kitaura F.-S.,  Hess S.,  2013, MNRAS, 435, 78,
  \eprint{1212.3514}

\bibitem[\protect\citeauthoryear{Kitaura, Jasche, Li, Ensslin, Metcalf
  et~al.,}{Kitaura et~al.}{2009}]{Kitaura:2009uh}
Kitaura F.~S.,  Jasche J.,  Li C.,  Ensslin T.~A.,  Metcalf R.~B.,    et~al.,
  2009, MNRAS, 400, 183, \eprint{0906.3978}

\bibitem[\protect\citeauthoryear{Kitaura, Yepes \& Prada}{Kitaura
  et~al.}{2014a}]{Kitaura:2013cwa}
Kitaura F.-S.,  Yepes G.,    Prada F.,  2014, MNRAS, 439, L21, \eprint{1307.3285}

\bibitem[\protect\citeauthoryear{Klypin \& Holtzman}{Klypin \&
  Holtzman}{1997}]{Klypin:1997sk}
Klypin A.,  Holtzman J.,  1997, \eprint{astro-ph/9712217}

\bibitem[\protect\citeauthoryear{Klypin, Yepes, Gottlober, Prada \&
  Hess}{Klypin et~al.}{2014}]{Klypin:2014kpa}
Klypin A.,  Yepes G.,  Gottlober S.,  Prada F.,    Hess S.,  2014,
  \eprint{1411.4001}

\bibitem[\protect\citeauthoryear{Kravtsov, Berlind, Wechsler, Klypin,
  Gottloeber et~al.,}{Kravtsov et~al.}{2004}]{Kravtsov:2003sg}
Kravtsov A.~V.,  Berlind A.~A.,  Wechsler R.~H.,  Klypin A.~A.,  Gottloeber S.,
     et~al., 2004, ApJ, 609, 35, \eprint{astro-ph/0308519}

\bibitem[\protect\citeauthoryear{Laureijs et~al.,}{Laureijs
  et~al.}{2011}]{Laureijs:2011gra}
Laureijs R.,  et~al., 2011, \eprint{1110.3193}

\bibitem[\protect\citeauthoryear{Levi et~al.,}{Levi
  et~al.}{2013}]{Levi:2013gra}
Levi M.,  et~al., 2013, \eprint{1308.0847}

\bibitem[\protect\citeauthoryear{McBride, Connolly, Gardner, Scranton, Newman
  et~al.,}{McBride et~al.}{2011}]{McBride:2010zn}
McBride C.~K.,  Connolly A.~J.,  Gardner J.~P.,  Scranton R.,  Newman J.~A.,
  et~al., 2011, ApJ, 726, 13, \eprint{1007.2414}

\bibitem[\protect\citeauthoryear{Manera, Samushia, Tojeiro, Howlett, Ross
  et~al.,}{Manera et~al.}{2015}]{Manera:2014cpa}
Manera M.,  Samushia L.,  Tojeiro R.,  Howlett C.,  Ross A.~J.,    et~al.,
  2015, MNRAS, 447, 437, \eprint{1401.4171}

\bibitem[\protect\citeauthoryear{Manera, Scoccimarro, Percival, Samushia,
  McBride et~al.,}{Manera et~al.}{2012}]{Manera:2012sc}
Manera M.,  Scoccimarro R.,  Percival W.~J.,  Samushia L.,  McBride C.~K.,
  et~al., 2012, MNRAS, 428, 1036, \eprint{1203.6609}

\bibitem[\protect\citeauthoryear{Monaco, Sefusatti, Borgani, Crocce, Fosalba
  et~al.,}{Monaco et~al.}{2013}]{Monaco:2013qta}
Monaco P.,  Sefusatti E.,  Borgani S.,  Crocce M.,  Fosalba P.,    et~al.,
  2013, MNRAS, 433, 2389, \eprint{1305.1505}

\bibitem[\protect\citeauthoryear{Monaco, Theuns \& Taffoni}{Monaco
  et~al.}{2002}]{Monaco:2001jg}
Monaco P.,  Theuns T.,    Taffoni G.,  2002, MNRAS, 331, 587,
  \eprint{astro-ph/0109323}

\bibitem[\protect\citeauthoryear{Nuza, Sanchez, Prada, Klypin, Schlegel
  et~al.,}{Nuza et~al.}{2013}]{Nuza:2012mw}
Nuza S.,  Sanchez A.,  Prada F.,  Klypin A.,  Schlegel D.,    et~al., 2013,
  MNRAS, 432, 743, \eprint{1202.6057}

\bibitem[\protect\citeauthoryear{Parkinson, Riemer-Sorensen, Blake, Poole,
  Davis et~al.,}{Parkinson et~al.}{2012}]{Parkinson:2012vd}
Parkinson D.,  Riemer-Sorensen S.,  Blake C.,  Poole G.~B.,  Davis T.~M.,
  et~al., 2012, Phys.Rev., D86, 103518, \eprint{1210.2130}

\bibitem[\protect\citeauthoryear{Percival et~al.,}{Percival
  et~al.}{2010}]{Percival:2009xn}
Percival W.~J.,  et~al., 2010, MNRAS, 401, 2148,
  \eprint{0907.1660}

\bibitem[\protect\citeauthoryear{Pope \& Szapudi}{Pope \&
  Szapudi}{2008}]{Pope:2007vz}
Pope A.~C.,  Szapudi I.,  2008, MNRAS, 389, 766,
  \eprint{0711.2509}

\bibitem[\protect\citeauthoryear{Reid, Percival, Eisenstein, Verde, Spergel
  et~al.,}{Reid et~al.}{2010}]{Reid:2009xm}
Reid B.~A.,  Percival W.~J.,  Eisenstein D.~J.,  Verde L.,  Spergel D.~N.,
  et~al., 2010, MNRAS, 404, 60, \eprint{0907.1659}

\bibitem[\protect\citeauthoryear{Riebe, Partl, Enke, Forero-Romero, Gottloeber
  et~al.,}{Riebe et~al.}{2011}]{Riebe:2011gp}
Riebe K.,  Partl A.~M.,  Enke H.,  Forero-Romero J.,  Gottloeber S.,    et~al.,
  2011, \eprint{1109.0003}

\bibitem[\protect\citeauthoryear{Samushia, Percival \& Raccanelli}{Samushia
  et~al.}{2012}]{Samushia:2011cs}
Samushia L.,  Percival W.~J.,    Raccanelli A.,  2012, MNRAS,
  420, 2102, \eprint{1102.1014}

\bibitem[\protect\citeauthoryear{Schlegel et~al.,}{Schlegel
  et~al.}{2011}]{Schlegel:2011zz}
Schlegel D.,  et~al., 2011, \eprint{1106.1706}

\bibitem[\protect\citeauthoryear{Scoccimarro \& Sheth}{Scoccimarro \&
  Sheth}{2002}]{Scoccimarro:2001cj}
Scoccimarro R.,  Sheth R.~K.,  2002, MNRAS, 329, 629,
  \eprint{astro-ph/0106120}

\bibitem[\protect\citeauthoryear{Skibba \& Sheth}{Skibba \&
  Sheth}{2009}]{Skibba:2008gr}
Skibba R.~A.,  Sheth R.~K.,  2009, MNRAS, 392, 1080,
  \eprint{0805.0310}

\bibitem[\protect\citeauthoryear{Springel}{Springel}{2005}]{Springel:2005mi}
Springel V.,  2005, MNRAS, 364, 1105,
  \eprint{astro-ph/0505010}

\bibitem[\protect\citeauthoryear{Tassev \& Zaldarriaga}{Tassev \&
  Zaldarriaga}{2012}]{Tassev:2011ac}
Tassev S.,  Zaldarriaga M.,  2012, JCAP, 1204, 013, \eprint{1109.4939}

\bibitem[\protect\citeauthoryear{Tassev, Zaldarriaga \& Eisenstein}{Tassev
  et~al.}{2013}]{Tassev:2013pn}
Tassev S.,  Zaldarriaga M.,    Eisenstein D.,  2013, JCAP, 1306, 036,
  \eprint{1301.0322}

\bibitem[\protect\citeauthoryear{Warren, Abazajian, Holz \& Teodoro}{Warren
  et~al.}{2006}]{Warren:2005ey}
Warren M.~S.,  Abazajian K.,  Holz D.~E.,    Teodoro L.,  2006, ApJ,
  646, 881, \eprint{astro-ph/0506395}

\bibitem[\protect\citeauthoryear{Watson, Iliev, D'Aloisio, Knebe, Shapiro
  et~al.,}{Watson et~al.}{2013}]{Watson:2012mt}
Watson W.~A.,  Iliev I.~T.,  D'Aloisio A.,  Knebe A.,  Shapiro P.~R.,
  et~al., 2013, MNRAS, 433, 1230, \eprint{1212.0095}

\bibitem[\protect\citeauthoryear{White}{White}{2014}]{White:2014gfa}
White M.,  2014, MNRAS, 439, 3630, \eprint{1401.5466}

\bibitem[\protect\citeauthoryear{White, Tinker \& McBride}{White
  et~al.}{2013}]{White:2013psd}
White M.,  Tinker J.~L.,    McBride C.~K.,  2013, \eprint{1309.5532}

\bibitem[\protect\citeauthoryear{Wild et~al.,}{Wild
  et~al.}{2005}]{Wild:2004me}
Wild V.,  et~al., 2005, MNRAS, 356, 247,
  \eprint{astro-ph/0404275}

\bibitem[\protect\citeauthoryear{York et~al.,}{York
  et~al.}{2000}]{York:2000gk}
York D.~G.,  et~al., 2000, AJ, 120, 1579, \eprint{astro-ph/0006396}

\bibitem[\protect\citeauthoryear{Zehavi et~al.,}{Zehavi
  et~al.}{2011}]{Zehavi:2010bh}
Zehavi I.,  et~al., 2011, ApJ, 736, 59, \eprint{1005.2413}

\bibitem[\protect\citeauthoryear{Zhao, Kitaura, Chuang, Prada, Yepes
  et~al.,}{Zhao et~al.}{2015}]{Zhao:2015jga}
Zhao C.,  Kitaura F.-S.,  Chuang C.-H.,  Prada F.,  Yepes G.,    et~al., 2015,
  \eprint{1501.05520}

\bibitem[\protect\citeauthoryear{Zheng, Berlind, Weinberg, Benson, Baugh
  et~al.,}{Zheng et~al.}{2005}]{Zheng:2004id}
Zheng Z.,  Berlind A.~A.,  Weinberg D.~H.,  Benson A.~J.,  Baugh C.~M.,
  et~al., 2005, ApJ, 633, 791, \eprint{astro-ph/0408564}

\end{thebibliography}
\end{document}